%% file: FSSCL.tex
\documentclass[journal,comsoc,10pt]{IEEEtran}

\usepackage{cite}

\usepackage{graphicx}
\usepackage{xcolor}
\usepackage{tikz}
\usetikzlibrary{arrows,positioning,shapes.geometric}
\usepackage{pgfplots}
\usepackage{multirow}
\usepackage{upgreek}
\usepackage{amssymb}
\usepackage{amsmath}
\usepackage{cases}
\usepackage{url}
\usepackage{pifont}
\usepackage{bm}
\usepackage{amsthm}
\newtheorem{theorem}{Theorem}

\newtheorem{remark}{Remark}
\usepackage{tabularx}
\usepackage{booktabs}
\usepackage{filecontents}
\usepackage{subcaption}
\newcolumntype{Y}{>{\centering\arraybackslash}X}

\usepackage{algorithmic}

\usepackage{array}

\newcommand{\fixme}[2]{\ifx&#2&{\leavevmode\color{red}#1}\else{\leavevmode\color{red}FIXME\{}#1{\leavevmode\color{red}\}}\footnote{{\leavevmode\color{red}#2}}\PackageWarning{Fixme}{#1: #2}\fi}

\newcommand{\newstuff}[2]{\ifx&#2&{\leavevmode\color{blue}#1}\else{\leavevmode\color{blue}FIXME\{}#1{\leavevmode\color{blue}\}}\footnote{{\leavevmode\color{blue}#2}}\PackageWarning{Newstuff}{#1: #2}\fi}

\DeclareMathOperator*{\argmin}{arg\,min}
\DeclareMathOperator*{\sgn}{sgn}
\DeclareMathOperator{\PM}{PM}
\DeclareMathOperator*{\arctanh}{arctanh}

\title{Fast and Flexible Successive-Cancellation List Decoders for Polar Codes}

\author{Seyyed~Ali~Hashemi,~\IEEEmembership{Student~Member,~IEEE,}
        Carlo~Condo,
        Warren~J.~Gross,~\IEEEmembership{Senior~Member,~IEEE}% <-this % stops a space
\thanks{This work has been published in parts in the IEEE Wireless Communications and Networking Conference Workshops (WCNCW), 2017 \cite{hashemi_FSSCL}.

S.~A.~Hashemi, C.~Condo, and W.~J.~Gross are with the Department of Electrical and Computer Engineering, McGill University, Montr\'eal, Qu\'ebec, Canada. e-mail: seyyed.hashemi@mail.mcgill.ca, carlo.condo@mail.mcgill.ca, warren.gross@mcgill.ca.}% <-this % stops a space
}

\begin{document}

\maketitle

\begin{abstract}
Polar codes have gained significant amount of attention during the past few years and have been selected as a coding scheme for the next generation of mobile broadband standard. Among decoding schemes, successive-cancellation list (SCL) decoding provides a reasonable trade-off between the error-correction performance and hardware implementation complexity when used to decode polar codes, at the cost of limited throughput.
The simplified SCL (SSCL) and its extension SSCL-SPC increase the speed of decoding by removing redundant calculations when encountering particular information and frozen bit patterns (rate one and single parity check codes), while keeping the error-correction performance unaltered.
In this paper, we improve SSCL and SSCL-SPC by proving that the list size imposes a specific number of path splitting required to decode rate one and single parity check codes. Thus, the number of splitting can be limited while guaranteeing exactly the same error-correction performance as if the paths were forked at each bit estimation. We call the new decoding algorithms Fast-SSCL and Fast-SSCL-SPC.
Moreover, we show that the number of path forks in a practical application can be tuned to achieve desirable speed, while keeping the error-correction performance almost unchanged. Hardware architectures implementing both algorithms are then described and implemented: it is shown that our design can achieve $\mathbf{1.86}$ Gb/s throughput, higher than the best state-of-the-art decoders.
\end{abstract}

\begin{IEEEkeywords}
polar codes, successive-cancellation decoding, list decoding, hardware implementation.
\end{IEEEkeywords}

\IEEEpeerreviewmaketitle

\section{Introduction} \label{sec:intro}

Polar codes are the first family of error-correcting codes with provable capacity-achieving property and a low-complexity encoding and decoding process \cite{arikan}. The successive-cancellation (SC) decoding is a low-complexity algorithm with which polar codes can achieve the capacity of a memoryless channel. However, there are two main drawbacks associated with SC. Firstly, SC requires the decoding process to advance bit by bit. This results in high latency and low throughput when implemented in hardware \cite{leroux}. Second, polar codes decoded with SC only achieve the channel capacity when the code length tends toward infinity. For practical polar codes of moderate length, SC falls short in providing a reasonable error-correction performance.

The first issue is a result of the serial nature of SC. In order to address this issue, the recursive structure of polar codes construction and the location of information and parity (frozen) bits were utilized in \cite{alamdar,sarkis} to identify constituent polar codes. In particular, rate zero (\mbox{Rate-0}) codes with all frozen bits, rate one (\mbox{Rate-1}) codes with all information bits, repetition (Rep) codes with a single information bit in the most reliable position, and single parity-check (SPC) codes with a single frozen bit in the least reliable position, were shown to be capable of being decoded in parallel with low-complexity decoding algorithms. This in turn increased the throughput and reduced the latency significantly. Moreover, the simplifications in \cite{alamdar,sarkis} did not introduce any error-correction performance degradation with respect to conventional SC.

The second issue stems from the fact that SC is suboptimal with respect to maximum-likelihood (ML) decoding. The decoding of each bit is only dependent on the bits already decoded. SC is unable to use the information about the bits that are not decoded yet. In order to address this issue, SC list (SCL) decoding advances by estimating each bit as either $0$ or $1$. Therefore, the number of candidate codewords doubles at each bit estimation step. In order to limit the exponential increase in the number of candidates, only $L$ candidate codewords are allowed to survive by employing a path metric ($\PM$) \cite{tal_list}. The $\PM$s were sorted and the $L$ best candidates were kept for further processing. It should be noted that SCL was previously used to decoder Reed-Muller codes \cite{Dumer_List}. SCL reduces the gap between SC and ML and it was shown that when a cyclic redundancy check (CRC) code is concatenated with polar codes, SCL can make polar codes outperform the state-of-the-art codes to the extent that polar codes have been chosen to be adopted in the next generation of mobile broadband standard \cite{3gpp_polar}.

The good error-correction performance of SCL comes at the cost of higher latency, lower throughput, and higher area occupation than SC when implemented on hardware \cite{balatsoukas_SCL_HW}. It was identified in \cite{balatsoukas} that using the log-likelihood ratio (LLR) values results in a SCL decoder which is more area-efficient than the conventional SCL decoder with log-likelihood (LL) values. In order to reduce the latency and increase the throughput associated with SCL, several attempts have been made to reduce the number of required decoding time steps as defined in \cite{arikan}. It should be noted that different time steps might entail different operations (e.g. a bit estimation or an LLR value update), and might thus last a different number of clock cycles. A group of $M$ bits were allowed to be decoded together in \cite{yuan_multibit_LLR,xiong_symbol}. \cite{lin_SCL} proposed a high throughput architecture based on a tree-pruning scheme and further extended it to a multimode decoder in \cite{xiong_multimode}. The throughput increase in \cite{lin_SCL} is based on code-based parameters which could degrade the error-correction performance significantly. Based on the idea in \cite{sarkis}, a fast list decoder architecture for software implementation was proposed in \cite{sarkis_list} which was able to decode constituent codes in a polar code in parallel. This resulted in fewer number of time steps to finish the decoding process. However, the SCL decoder in \cite{sarkis_list} is based on an empirical approach to decode constituent \mbox{Rate-1} and SPC codes and cannot guarantee the same error-correction performance as the conventional SCL decoder. Moreover, all the decoders in \cite{lin_SCL,xiong_multimode,sarkis_list} require a large sorter to select the surviving candidate codewords. Since the sorter in the hardware implementation of SCL decoders has a long and dominant critical path which is dependent on the number of its inputs \cite{balatsoukas}, increasing the number of $\PM$s results in a longer critical path and a lower operating frequency.

Based on the idea of list sphere decoding in \cite{hashemi_LSD}, a simplified SCL (SSCL) was proposed in \cite{hashemi_SSCL} which identified and avoided the redundant calculations in SCL. Therefore, it required fewer number of time steps than SCL to decode a polar code. The advantage of SSCL is that it not only guarantees the error-correction performance preservation, but also it uses the same sorter as in the conventional SCL algorithm. To further increase the throughput and reduce the latency of SSCL, the matrix reordering idea in \cite{hashemi_MR} was used to develop the SSCL-SPC decoder in \cite{hashemi_SSCL_TCASI}. While SSCL-SPC uses the same sorter as in the conventional SCL, it provides an exact reformulation for $L=2$ and its approximations bring negligible error-correction performance loss with respect to SSCL.

While SSCL and SSCL-SPC are algorithms that can work with any list size, they fail to address the redundant path splitting associated with a specific list size. In this paper, we first prove that there is a specific number of path splitting required for decoding the constituent codes in SSCL and SSCL-SPC for every list size to guarantee the error-correction performance preservation. Any path splitting after that number is redundant and any path splitting before that number cannot provably preserve the error-correction performance. Since these decoders require fewer number of time steps than SSCL and SSCL-SPC, we name them Fast-SSCL and Fast-SSCL-SPC, respectively. We further show that in practical polar codes, we can achieve similar error-correction performance to SSCL and SSCL-SPC with even fewer number of path forks. Therefore, we can optimize Fast-SSCL and Fast-SSCL-SPC for speed. We propose hardware architectures to implement both new algorithms: implementation results yield the highest throughput in the state-of-the-art with comparable area occupation.

This paper is an extension to our work in \cite{hashemi_FSSCL} in which the Fast-SSCL algorithm was proposed. Here, we propose the Fast-SSCL-SPC algorithm and prove that its error-correction performance is identical to that of SSCL-SPC. We further propose speed-up techniques for Fast-SSCL and Fast-SSCL-SPC which incur almost no error-correction performance loss. Finally, we propose hardware architectures implementing the aforementioned algorithms and show the effectiveness of the proposed techniques by comparing our designs with state of the art.

The remainder of this paper is organized as follows: Section~\ref{sec:prel} provides a background on polar codes and its decoding algorithms. Section~\ref{sec:FSSCL} introduces the proposed Fast-SSCL and Fast-SSCL-SPC algorithms and their speed optimization technique. A decoder architecture is proposed in Section~\ref{sec:decarch} and the implementation results are provided in Section~\ref{sec:results}. Finally, Section~\ref{sec:conc} draws the main conclusions of the paper.

\section{Preliminaries} 
\label{sec:prel}

\subsection{Polar Codes}\label{sec:prel:polar}

A polar code of length $N$ with $K$ information bits is represented by $\mathcal{P}(N,K)$ and can be constructed recursively with two polar codes of length $N/2$. The encoding process can be denoted as a matrix multiplication as $\mathbf{x} = \mathbf{u}\mathbf{G}_N$, where $\mathbf{u} = \{u_0,u_1,\ldots,u_{N-1}\}$ is the sequence of input bits, $\mathbf{x} = \{x_0,x_1,\ldots,x_{N-1}\}$ is the sequence of coded bits, and $\mathbf{G}_N = \mathbf{B}_N\mathbf{G}^{\otimes n}$ is the generator matrix created by the product of $\mathbf{B}_N$ which is the bit-reversal permutation matrix, and $\mathbf{G}^{\otimes n}$ which is the $n$-th Kronecker product of the polarizing matrix $\mathbf{G}=\left[\begin{smallmatrix} 1&0\\ 1&1 \end{smallmatrix} \right]$.

The encoding process involves the determination of the $K$ bit-channels with the best channel characteristics and assigning the information bits to them. The remaining $N-K$ bit-channels are set to a known value known at the decoder side. They are thus called frozen bits with set $\mathcal{F}$. Since the value of these bits does not have an impact on the error-correction performance of polar codes on a symmetric channel, they are usually set to $0$. The codeword $\mathbf{x}$ is then modulated and sent through the channel. In this paper, we consider binary phase-shift keying (BPSK) modulation which maps $\{0,1\}$ to $\{+1,-1\}$.

\subsection{Successive-Cancellation Decoding}
\label{sec:prel:SCDec}

\begin{figure}
  \centering
  \input{figures/sc-dec.tikz}
  \caption{SC decoding on a binary tree for $\mathcal{P}(8,4)$ and $\{u_0,u_1,u_2,u_4\}\in \mathcal{F}$.}
  \label{figSCDec}
\end{figure}
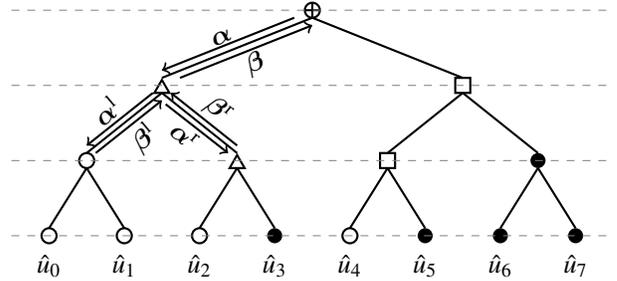

The SC decoding process can be represented as a binary tree search as shown in Fig.~\ref{figSCDec} for $\mathcal{P}(8,4)$. Thanks to the recursive construction of polar codes, at each stage $s$ of the tree, each node can be interpreted as a polar code of length $N_s = 2^s$. Two kinds of messages are passed between the nodes, namely, soft LLR values $\bm{\alpha} = \{\alpha_0,\alpha_1,\ldots,\alpha_{N_s-1}\}$ which are passed from parent to child nodes, and the hard bit estimates $\bm{\beta} = \{\beta_0,\beta_1,\ldots,\beta_{N_s-1}\}$ which are passed from child nodes to the parent node.

The $\frac{N_s}{2}$ elements of the left child node $\bm{\alpha}^\text{l} = \{\alpha^\text{l}_0,\alpha^\text{l}_1,\ldots,\alpha^\text{l}_{\frac{N_s}{2}-1}\}$, and the right child node $\bm{\alpha}^\text{r} = \{\alpha^\text{r}_0,\alpha^\text{r}_1,\ldots,\alpha^\text{r}_{\frac{N_s}{2}-1}\}$, can be computed as \cite{arikan}
\begin{align}
\alpha^\text{l}_i =& 2\arctanh \left(\tanh\left(\frac{\alpha_i}{2}\right)\tanh\left(\frac{\alpha_{i+\frac{N_s}{2}}}{2}\right)\right) \text{,} \label{eq1} \\
\alpha^{\text{r}}_i =& \alpha_{i+\frac{N_s}{2}} + \left(1-2\beta^\text{l}_i\right)\alpha_i \text{,}
\label{eq2}
\end{align}
whereas the $N_s$ values of $\bm{\beta}$ are calculated by means of the left child and right child node messages $\bm{\beta}^\text{l} = \{\beta^\text{l}_0,\beta^\text{l}_1,\ldots,\beta^\text{l}_{\frac{N_s}{2}-1}\}$ and $\bm{\beta}^\text{r} = \{\beta^\text{r}_0,\beta^\text{r}_1,\ldots,\beta^\text{r}_{\frac{N_s}{2}-1}\}$ as \cite{arikan}
\begin{equation}
\beta_i =
  \begin{cases}
    \beta^\text{l}_i\oplus \beta^\text{r}_i \text{,} & \text{if} \quad i < \frac{N_s}{2} \text{,}\\
    \beta^\text{r}_{i-\frac{N_s}{2}} \text{,} & \text{otherwise} \text{,}
  \end{cases}
  \label{eq3}
\end{equation}
where $\oplus$ is the bitwise XOR operation. At leaf nodes, the $i$-th bit $\hat{u}_i$ can be estimated as
\begin{equation}
\hat{u}_i =
  \begin{cases}
    0 \text{,} & \text{if } i \in \mathcal{F} \text{ or } \alpha_{i}\geq 0\text{,}\\
    1 \text{,} & \text{otherwise.}
  \end{cases} \label{eq6}
\end{equation}
Equation~(\ref{eq1}) can be reformulated in a more hardware-friendly (HWF) version that has first been proposed in \cite{leroux}:
\begin{equation}
\alpha^{\text{l}}_i = \sgn(\alpha_i)\sgn(\alpha_{i+\frac{N_s}{2}})\min(|\alpha_i|,|\alpha_{i+\frac{N_s}{2}}|) \text{.} \label{eq4}
\end{equation}

\subsection{Successive-Cancellation List Decoding} \label{sec:prel:SCLDec}

The error-correction performance of SC when applied to codes with short to moderate length can be improved by the use of SCL-based decoding. The SCL algorithm estimates a bit considering both its possible values $0$ and $1$. At every estimation, the number of codeword candidates (paths) doubles: in order to limit the increase in the complexity of this algorithm, only a set of $L$ codeword candidates is memorized at all times. Thus, after every estimation, half of the paths are discarded. To this purpose, a $\PM$ is associated to each path and updated at every new estimation: it can be considered a cost function, and the $L$ paths with the lowest $\PM$s are allowed to survive. In the LLR-based SCL \cite{balatsoukas}, the $\PM$ can be computed as
\begin{equation} 
\PM_{i_l} = \sum_{j = 0}^i \ln\left(1+\mathrm{e}^{-(1-2\hat{u}_{j_l})\alpha_{j_l}}\right) \text{,} \label{eq5}
\end{equation}
where $l$ is the path index and $\hat{u}_{j_l}$ is the estimate of bit $j$ at path $l$. A HWF version of Equation~(\ref{eq5}) has been proposed in \cite{balatsoukas}:
\begin{align}
&\PM_{{-1}_l} = 0 \text{,} \nonumber \\
&\PM_{{i}_l} = \begin{cases}
    \PM_{{i-1}_l} + |\alpha_{i_l}| \text{,} & \text{if } \hat{u}_{i_l} \neq \frac{1}{2}\left(1-\sgn\left(\alpha_{i_l}\right)\right)\text{,}\\
    \PM_{{i-1}_l} \text{,} & \text{otherwise,}
  \end{cases} \label{eq7}
\end{align}
which can be rewritten as
\begin{equation}
\PM_{{i}_l} = \frac{1}{2}\sum_{j = 0}^{i}\sgn(\alpha_{{{j}_l}})\alpha_{{{j}_l}} - (1-2\hat{u}_{j_l})\alpha_{{{j}_l}} \text{.} \label{eq7_1}   
\end{equation}

In case the hardware does not introduce bottlenecks and both (\ref{eq2}) and (\ref{eq4}) can be computed in a single time step, the number of time steps required to decode a code of length $N$ with $K$ information bits in SCL is \cite{balatsoukas}
\begin{equation}
T_{\text{SCL}}(N,K) = 2N + K - 2 \text{.}
\end{equation}

\subsection{Simplified Successive-Cancellation List Decoding} \label{sec:prel:SSCLDec}

\subsubsection{SSCL Decoding}

The SSCL algorithm in \cite{hashemi_SSCL} provides efficient decoders for \mbox{Rate-0}, Rep, and \mbox{Rate-1} nodes in SCL without traversing the decoding tree while guaranteeing the error-correction performance preservation. For example in Fig.~\ref{figSCDec}, the black circles represent \mbox{Rate-1} nodes, the white circles represent \mbox{Rate-0} nodes, and the white triangles represent Rep nodes. The pruned decoding tree of SSCL for the example in Fig.~\ref{figSCDec} is shown in Fig.~\ref{figSSCLDec} which consists of two Rep nodes and a \mbox{Rate-1} node.

Let us consider that the vectors $\bm{\alpha}_l$ and $\bm{\eta}_l = 1-2\bm{\beta}_l$ are relative to the top of a node in the decoding tree. \mbox{Rate-0} nodes can be decoded as
\begin{subnumcases}{\kern-2em\PM_{{N_s-1}_l}=}
\sum_{i = 0}^{N_s-1} \ln\left(1+\mathrm{e}^{-\alpha_{i_l}}\right) \text{,} & \kern-1em\text{Exact,} \label{eq:Rate0:Exact} \\ \frac{1}{2} \sum_{i = 0}^{N_s-1} \sgn\left(\alpha_{i_l}\right)\alpha_{i_l} - \alpha_{i_l} \text{,} & \kern-1em\text{HWF.} \label{eq:Rate0:HWF}
\end{subnumcases}
Rep nodes can be decoded as
\begin{subnumcases}{\kern-2em\PM_{{N_s-1}_l}=}
\sum_{i = 0}^{N_s-1} \ln\left(1+\mathrm{e}^{-\eta_{{N_s-1}_l}\alpha_{i_l}}\right) \text{,} & \kern-1em\text{Exact,} \label{eq:Rep:Exact} \\
\frac{1}{2} \sum_{i = 0}^{N_s-1} \sgn\left(\alpha_{i_l}\right)\alpha_{i_l} - \eta_{{N_s-1}_l}\alpha_{i_l} \text{,} & \kern-1em\text{HWF.} \label{eq:Rep:HWF}
\end{subnumcases}
where $\eta_{{N_s-1}_l}$ represents the bit estimate of the information bit in the Rep node. Finally, \mbox{Rate-1} nodes can be decoded as
\begin{subnumcases}{\kern-2em\PM_{{N_s-1}_l}=}
\sum_{i = 0}^{N_s-1} \ln\left(1+\mathrm{e}^{-\eta_{i_l}\alpha_{i_l}}\right) \text{,} & \kern-1em\text{Exact,} \label{eq:Rate1:Exact} \\
\frac{1}{2} \sum_{i = 0}^{N_s-1} \sgn\left(\alpha_{i_l}\right)\alpha_{i_l} - \eta_{i_l}\alpha_{i_l} \text{,} & \kern-1em\text{HWF.} \label{eq:Rate1:HWF}
\end{subnumcases}

It was shown in \cite{hashemi_SSCL_TCASI} that the time step requirements of \mbox{Rate-0}, Rep, and \mbox{Rate-1} nodes of length $N_s$ in SSCL decoding can be represented as
\begin{align}
T_{\text{SSCL}_{\text{Rate-0}}}(N_s,0) &= 1 \text{,} \\
T_{\text{SSCL}_{\text{Rep}}}(N_s,1) &= 2 \text{,} \\
T_{\text{SSCL}_{\text{Rate-1}}}(N_s,N_s) &= N_s \text{.}
\end{align}

While the SSCL algorithm reduces the number of required time steps to decode \mbox{Rate-1} nodes by almost a factor of three, it fails to address the effect of list size on the maximum number of required path forks. In Section \ref{sec:FSSCL}, we prove that the number of required time steps to decode \mbox{Rate-1} nodes depends on the list size and that the new \mbox{Fast-SSCL} algorithm is faster than both SCL and SSCL without incurring any error-correction performance degradation.

\subsubsection{SSCL-SPC Decoding}

\begin{figure}
  \centering
\begin{subfigure}{0.24\textwidth}
  \centering
  \input{figures/sscl-dec.tikz}
  \caption{SSCL}
  \label{figSSCLDec}
\end{subfigure}
\begin{subfigure}{0.24\textwidth}
  \centering
  \input{figures/ssclspc-dec.tikz}
  \caption{SSCL-SPC}
  \label{figSSCLSPCDec}
\end{subfigure}
  \caption{(a) SSCL, and (b) SSCL-SPC decoding tree for $\mathcal{P}(8,4)$ and $\{u_0,u_1,u_2,u_4\}\in \mathcal{F}$.}
  \label{figSSCL_SSCLSPC}
\end{figure}
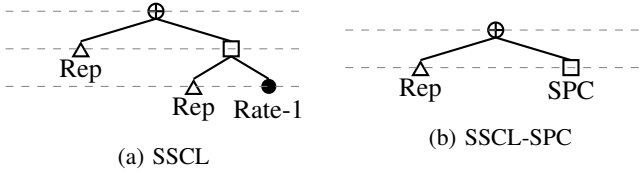

In \cite{hashemi_SSCL_TCASI}, a low-complexity approach was proposed to decode SPC nodes which resulted in exact reformulations for $L=2$ and its approximations for other list sizes brought negligible error-correction performance degradation. The pruned tree of \mbox{SSCL-SPC} for the same example as in Fig.~\ref{figSCDec} is shown in Fig.~\ref{figSSCLSPCDec} which consists of a Rep node and a SPC node. The idea is to decode the frozen bit in SPC nodes in the first step of the decoding process. In order to do that, the $\PM$ calculations in the HWF formulation were carried out by only finding the LLR value of the least reliable bit and using the LLR values at the top of the polar code tree in the SCL decoding algorithm for the rest of the bits.

The least reliable bit in an SPC node of length $N_s$ is found as
\begin{equation} \label{eq:SPCmin}
i_{\min} = \underset{0\leq i <N_s}{\argmin}(|\alpha_i|) \text{,}
\end{equation}
and the parity of it is derived as
\begin{equation}
\gamma = \bigoplus_{i=0}^{N_s-1} \left(\frac{1}{2}\left(1-\sgn\left(\alpha_i\right)\right)\right) \text{.} \label{eq:SPCgamma}
\end{equation}
To satisfy the even-parity constraint, $\gamma$ is found for each path based on (\ref{eq:SPCgamma}). The $\PM$s are then initialized as
\begin{equation} \label{eq:SPCPM0}
\PM_0 =
  \begin{cases}
    \PM_{-1} + |\alpha_{i_{\min}}| \text{,} & \text{if } \gamma = 1 \text{,} \\
    \PM_{-1} \text{,} & \text{otherwise.}
  \end{cases}
\end{equation}
In this way, the least reliable bit which corresponds to the even-parity constraint is decoded first. For bits other than the least reliable bit, the $\PM$ is updated as
\begin{equation} \label{eq:SPCPM}
\PM_i =
  \begin{cases}
    \PM_{i-1} + |\alpha_i| + (1-2\gamma)|\alpha_{i_{\min}}| \text{,} & \text{if } \eta_i \neq \sgn\left(\alpha_i\right) \text{,} \\
    \PM_{i-1} \text{,} & \text{otherwise.}
  \end{cases}
\end{equation}
Finally, when all the bits are estimated, the least reliable bit is set to preserve the even-parity constraint as
\begin{equation}
\beta_{i_{\min}} = \bigoplus_{\substack{i = 0\\i \neq i_{\min}}}^{N_s-1} \beta_i \text{.} \label{eq:SPCbitCorrection}
\end{equation}

In \cite{hashemi_SSCL_TCASI}, the time step requirements of SPC nodes of length $N_s$ in SSCL-SPC decoding was shown to be
\begin{equation}
T_{\text{SSCL-SPC}_{\text{SPC}}}(N_s,N_s-1) = N_s+1 \text{,}
\end{equation}
which consists of one time step for (\ref{eq:SPCPM0}), $N_s-1$ time steps for (\ref{eq:SPCPM}), and one time step for (\ref{eq:SPCbitCorrection}).

The SSCL-SPC algorithm reduces the number of required time steps to decode SPC nodes by almost a factor of three, but as in the case of \mbox{Rate-1} nodes, it fails to address the effect of list size on the maximum number of required path forks. In Section \ref{sec:FSSCL}, we prove that the number of required time steps to decode SPC nodes depends on the list size and that the new \mbox{Fast-SSCL-SPC} algorithm is faster than SSCL-SPC without incurring any error-correction performance degradation.

\section{Fast-SSCL Decoding}\label{sec:FSSCL}

In this section, we propose a fast decoding approach for \mbox{Rate-1} nodes and use it to develop \mbox{Fast-SSCL}. We further propose a fast decoding approach for SPC nodes in \mbox{SSCL-SPC} and use it to develop \mbox{Fast-SSCL-SPC}. To this end, we provide the exact number of path forks in \mbox{Rate-1} and SPC nodes to guarantee error-correction performance preservation. Any path splitting after that number is redundant and any path splitting less than that number cannot guarantee the error-correction performance preservation. We further show that in practical applications, this number can be reduced with almost no error-correction performance loss. We use this phenomenon to optimize Fast-SSCL and Fast-SSCL-SPC for speed.

\subsection{Guaranteed Error-Correction Performance Preservation} \label{sec:FSSCL:theorems}

The fast \mbox{Rate-1} and SPC decoders can be summarized by the following theorems.

\begin{theorem} \label{th:maxEstimate}
In SSCL decoding with list size $L$, the number of path splitting in a \mbox{Rate-1} node of length $N_s$ required to get the exact same results as the conventional SSCL decoder is
\begin{equation}
\min \left(L-1,N_s\right) \label{eq:maxEstimate} \text{.}
\end{equation}
\end{theorem}

The proposed technique results in $T_{\text{Fast-SSCL}_{\text{Rate-1}}}(N_s,N_s) = \min \left(L-1,N_s\right)$ which improves the required number of time steps to decode \mbox{Rate-1} nodes when $L-1<N_s$. Every bit after the $L-1$-th can be obtained through hard decision on the LLR as
\begin{equation}
\beta_{i_l} =
  \begin{cases}
    0 \text{,} & \text{if } \alpha_{i_l}\geq 0\text{,}\\
    1 \text{,} & \text{otherwise,}
  \end{cases} \label{eq:hardDecLLR}
\end{equation}
without the need for path splitting. On the other hand, in case $\min \left(L-1,N_s\right)=N_s$, all bits of the node need to be estimated and the decoding automatically reverts to the process described in \cite{hashemi_SSCL}. The proof of the theorem is nevertheless valid for both $L-1<N_s$ and $L-1\geq N_s$ and is provided in \cite{hashemi_FSSCL}.

The proposed theorem remains valid also for the HWF formulation that can be written as
\begin{equation}
\PM_{{i}_l} = \begin{cases}
    \PM_{{i-1}_l} + |\alpha_{i_l}|, & \text{if } \eta_{i_l} \neq \sgn\left(\alpha_{i_l}\right)\text{,}\\
    \PM_{{i-1}_l}, & \text{otherwise.}
  \end{cases} \label{eq:pmUpdateHW}
\end{equation}
The proof of the theorem in the HWF formulation case is also presented in \cite{hashemi_FSSCL}.

The result of Theorem~\ref{th:maxEstimate} provides an exact number of path forks in \mbox{Rate-1} nodes for each list size in SCL decoding in order to guarantee error-correction performance preservation. The \mbox{Rate-1} node decoder of \cite{sarkis_list} empirically states that two path forks are required to preserve the error-correction performance. The following remarks are the direct results of Theorem~\ref{th:maxEstimate}.

\begin{remark}
The \mbox{Rate-1} node decoder of \cite{sarkis_list} for $L=2$ is redundant.
\end{remark}
Theorem~\ref{th:maxEstimate} states that for a \mbox{Rate-1} node of length $N_s$ when $L=2$, the number of path splitting is $\min(L-1,N_s) = 1$. Therefore, there is no need to split the path after the least reliable bit is estimated. \cite{sarkis_list} for $L=2$ is thus redundant.

\begin{figure}
  \centering
  \hspace{-15pt}
  \input{figures/fer1kL128_1024_860.tikz}
  \input{figures/ber1kL128_1024_860.tikz}
  \\
  \hspace{20pt}\ref{perf-legend1kL128_1024_860}\vspace{2pt}
  \caption{FER and BER performance comparison of SSCL \cite{hashemi_SSCL} and the empirical method of \cite{sarkis_list} for $\mathcal{P}(1024,860)$ when $L=128$. The CRC length is $32$.}
  \label{fig:ER1kL128_1024_860}
\end{figure}
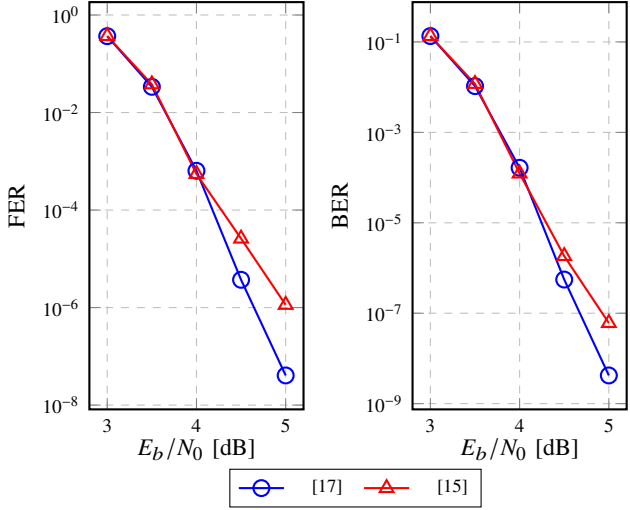

\begin{remark}
The \mbox{Rate-1} node decoder of \cite{sarkis_list} falls short in preserving the error-correction performance for higher rates and larger list sizes.
\end{remark}
For codes of higher rates, the number of \mbox{Rate-1} nodes of larger length increases \cite{hashemi_SSCL_TCASI}. Therefore, when the list size is also large, $\min(L-1,N_s) \gg 2$. The gap between the empirical method of \cite{sarkis_list} and the result of Theorem~\ref{th:maxEstimate} can introduce significant error-correction performance loss. Fig.~\ref{fig:ER1kL128_1024_860} provides the frame error rate (FER) and bit error rate (BER) of decoding a $\mathcal{P}(1024,860)$ code with SSCL of \cite{hashemi_SSCL} and the empirical method of \cite{sarkis_list} when the list size is $128$. It can be seen that the error-correction performance loss reaches $0.25$dB at FER of $10^{-5}$. In Section~\ref{sec:FSSCL:speed}, we show that the number of path forks can be tuned for each list size to find a good trade-off between the error-correction performance and the speed of decoding.

\begin{theorem} \label{th:maxEstimateSPC}
In \mbox{SSCL-SPC} decoding with list size $L$, the number of path forks in a SPC node of length $N_s$ required to get the exact same results as the conventional \mbox{SSCL-SPC} decoder is
\begin{equation}
\min \left(L,N_s\right) \label{eq:maxEstimateSPC} \text{.}
\end{equation}
\end{theorem}

Following the time step calculation of SSCL-SPC, the proposed technique in Theorem~\ref{th:maxEstimateSPC} results in $T_{\text{Fast-SSCL-SPC}_{\text{SPC}}}(N_s,N_s-1) = \min \left(L,N_s\right) + 1$ which improves the required number of time steps to decode SPC nodes when $L<N_s$. Every bit after the $L$-th can be obtained through hard decision on the LLR as in (\ref{eq:hardDecLLR}) without the need for path splitting. In case $\min \left(L,N_s\right)=N_s$, the paths need to be split for all bits of the node and the decoding automatically reverts to the process described in \cite{hashemi_SSCL_TCASI}. The proof of the theorem is nevertheless valid for both $L<N_s$ and $L\geq N_s$. We defer the proof to Appendix~\ref{sec:app2}.

The effectiveness of hard decision decoding after the $\min(L-1,N_s)$-th bit in Rate-1 nodes and the $\min(L,N_s)$-th bit in SPC nodes is due to the fact that the bits with high absolute LLR values are more reliable and less likely to incur path splitting. However, whether path splitting must occur or not depends on the list size $L$. The proposed \mbox{Rate-1} node decoder is used in \mbox{Fast-SSCL} and \mbox{Fast-SSCL-SPC} algorithms and the proposed SPC node decoder is used in \mbox{Fast-SSCL-SPC}, while the decoders for \mbox{Rate-0} and Rep nodes remain similar to those used in SSCL \cite{hashemi_SSCL} such that
\begin{align}
T_{\text{Fast-SSCL}_\text{Rate-0}}(N_s,0) &= T_{\text{Fast-SSCL-SPC}_\text{Rate-0}}(N_s,0) = 1 \text{,} \\
T_{\text{Fast-SSCL}_\text{Rep}}(N_s,1) &= T_{\text{Fast-SSCL-SPC}_\text{Rep}}(N_s,1) = 2 \text{.}
\end{align}

It should be noted that the number of path forks is directly related to the number of time steps required in the decoding process \cite{balatsoukas}. Therefore, when $L < N_s$, the time step requirement of SPC nodes based on Theorem~\ref{th:maxEstimateSPC} is two time steps more than the time step requirement of \mbox{Rate-1} nodes as in Theorem~\ref{th:maxEstimate}. However, if SPC nodes are not taken into account as in Fast-SSCL decoding, the polar code tree needs to be traversed to find Rep nodes and Rate-1 nodes as shown in Fig.~\ref{figSSCLDec}. For a SPC node of length $N_s$, this will result in additional time step requirements as
\begin{align*}
T_{\text{Fast-SSCL}_\text{SPC}}(N_s,N_s-1) =& 2 \log_2 N_s - 2 + T_{\text{Fast-SSCL}_{\text{Rep}}}(2,1) \nonumber \\
&+ \sum_{i = 1}^{\log_2 N_s - 1} T_{\text{Fast-SSCL}_{\text{Rate-1}}}(2^i,2^i) \text{.}
\end{align*}
For example, for a SPC node of length $64$, Fast-SSCL with $L = 4$ results in $T_{\text{Fast-SSCL}_\text{SPC}}(64,63) = 26$, while Fast-SSCL-SPC with $L = 4$ results in $T_{\text{Fast-SSCL-SPC}_\text{SPC}}(64,63) = 5$. Table~\ref{tab:timesteps} summarizes the number of time steps required to decode each node with different decoding algorithms.

\begin{table*}
	\centering
	\caption{Time-Step Requirements of Decoding Different Nodes of Length $N_s$ with List Size $L$.}
	\label{tab:timesteps}
		\setlength{\extrarowheight}{1.7pt}
%\scriptsize
	\begin{tabularx}{.67\textwidth}{lcccc}
\toprule

Algorithm & Rate-0 & Rep & Rate-1 & SPC \\

\cmidrule(lr){1-1}
\cmidrule(lr){2-2}
\cmidrule(lr){3-3}
\cmidrule(lr){4-4}
\cmidrule(lr){5-5}

SCL & $2N_s-2$ & $2N_s-1$ & $3N_s-2$ & $3N_s-3$ \\
SSCL & $1$ & $2$ & $N_s$ & $N_s + 2\log_2 N_s - 2$ \\
SSCL-SPC & $1$ & $2$ & $N_s$ & $N_s + 1$ \\
Fast-SSCL & $1$ & $2$ & $\min(L-1,N_s)$ & $2\log_2 N_s + \sum_{i = 1}^{\log_2 N_s - 1} \min(L-1,\frac{N_s}{2^i})$ \\
Fast-SSCL-SPC & $1$ & $2$ & $\min(L-1,N_s)$ & $\min(L,N_s) + 1$ \\

\bottomrule
	\end{tabularx}
\end{table*}

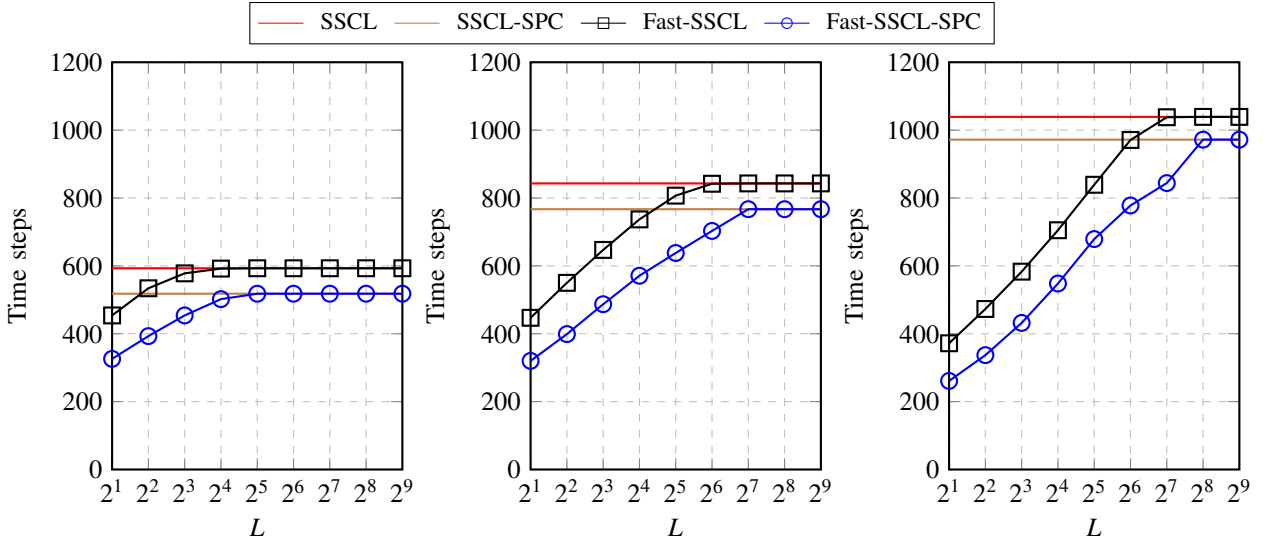
\begin{figure*}
  \centering
  \ref{perf-legendTimeReq}\\
\begin{subfigure}{0.3\textwidth}
  \input{figures/time-req025.tikz}
  \caption{$\mathcal{P}(1024,256)$}
  \label{figTimeReq25}
\end{subfigure}
\begin{subfigure}{0.3\textwidth}
  \input{figures/time-req05.tikz}
  \caption{$\mathcal{P}(1024,512)$}
  \label{figTimeReq50}
\end{subfigure}
\begin{subfigure}{0.3\textwidth}
  \input{figures/time-req075.tikz}
  \caption{$\mathcal{P}(1024,768)$}
  \label{figTimeReq75}
\end{subfigure}
  \caption{Time-step requirements of SSCL, SSCL-SPC, Fast-SSCL, and Fast-SSCL-SPC decoding of (a) $\mathcal{P}(1024,256)$, (b) $\mathcal{P}(1024,512)$, and (c) $\mathcal{P}(1024,768)$.}
  \label{figTimeReq}
\end{figure*}

In practical polar codes, there are many instances where $L-1<N_s$ for \mbox{Rate-1} nodes and using the \mbox{Fast-SSCL} algorithm can significantly reduce the number of required decoding time steps with respect to SSCL. Similarly, there are many instances where $L<N_s$ for \mbox{SPC} nodes and using the \mbox{Fast-SSCL-SPC} algorithm can significantly reduce the number of required decoding time steps with respect to SSCL-SPC. Fig.~\ref{figTimeReq} shows the savings in time step requirements of a polar code with three different rates. It should be noted that as the rate increases, the number of \mbox{Rate-1} and SPC nodes increases. This consequently results in more savings by going from SSCL (SSCL-SPC) to Fast-SSCL (Fast-SSCL-SPC).

\subsection{Speed Optimization} \label{sec:FSSCL:speed}

The analysis in Section~\ref{sec:FSSCL:theorems} provides exact reformulations of SSCL and SSCL-SPC decoders without introducing any error-correction performance loss. However, in practical polar codes, there are fewer required path forks for Fast-SSCL and Fast-SSCL-SPC in order to match the error-correction performance of SSCL and SSCL-SPC, respectively.

Without loss of generality, let us consider $L-1 < N_s$ for \mbox{Rate-1} nodes and $L < N_s$ for SPC nodes such that Fast-SSCL and Fast-SSCL-SPC result in higher decoding speeds than SSCL and SSCL-SPC, respectively. Let us now consider $S_{\text{Rate-1}}$ be the number of path forks in a \mbox{Rate-1} node of length $N_s$, and $S_{\text{SPC}}$ be the number of path forks in a SPC node of length $N_s$ where $S_{\text{Rate-1}} \leq L-1$ and $S_{\text{SPC}} \leq L$. It should be noted that $S_{\text{Rate-1}} = L-1$ and $S_{\text{SPC}} = L$ result in optimal number of path forks as presented in Theorem~\ref{th:maxEstimate} and Theorem~\ref{th:maxEstimateSPC}, respectively. The smaller the values of $S_{\text{Rate-1}}$ and $S_{\text{SPC}}$, the faster the decoders of Fast-SSCL and Fast-SSCL-SPC. Similar to (\ref{eq:maxEstimate}) and (\ref{eq:maxEstimateSPC}), the new number of required path forks for \mbox{Rate-1} and SPC nodes can be stated as $\min(S_\text{Rate-1},N_s)$ and $\min(S_\text{SPC},N_s)$, respectively.

\begin{figure}
  \centering
  \hspace{-15pt}
  \input{figures/fer1kL8SPC.tikz}
  \input{figures/ber1kL8SPC.tikz}
  \\
  \hspace{20pt}\ref{perf-legend1kL8SPC}\vspace{2pt}
  \caption{FER and BER performance comparison of Fast-SSCL-SPC decoding of $\mathcal{P}(1024,512)$ for $L=8$ and different values of $S_\text{Rate-1}$ and $S_\text{SPC}$. The CRC length is $16$.}
  \label{fig:ER1kL8SPC}
\end{figure}
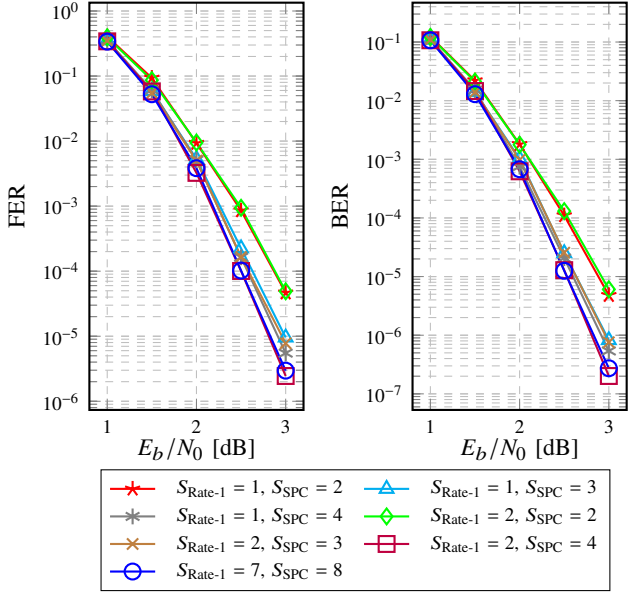

The definition of the parameters $S_{\text{Rate-1}}$ and $S_{\text{SPC}}$ provides a trade-off between error-correction performance and speed of Fast-SSCL and Fast-SSCL-SPC. Let us consider CRC-aided Fast-SSCL decoding of $\mathcal{P}(1024,512)$ with CRC length $16$. Fig.~\ref{fig:ER1kL2} shows that for $L=2$, choosing $S_{\text{Rate-1}} = 0$ results in significant FER and BER error-correction performance degradation. Therefore, when $L=2$, the optimal value of $S_{\text{Rate-1}} = 1$ is used for Fast-SSCL. The optimal value of $S_{\text{Rate-1}}$ for $L=4$ is $3$. However, as shown in Fig.~\ref{fig:ER1kL4}, $S_{\text{Rate-1}} = 1$ results in almost the same FER and BER performance as the optimal value of $S_{\text{Rate-1}} = 3$. For $L=8$, the selection of $S_{\text{Rate-1}} = 1$ results in ${\sim}0.1$~dB of error-correction performance degradation at FER~$=10^{-5}$ as shown in Fig.~\ref{fig:ER1kL8}. However, selecting $S_{\text{Rate-1}} = 2$ removes the error-correction performance gap to the optimal value of $S_{\text{Rate-1}} = 7$. In the case of CRC-aided Fast-SSCL-SPC decoding of $\mathcal{P}(1024,512)$ with $16$ bits of CRC, selecting $S_{\text{Rate-1}} = 1$ and $S_{\text{SPC}} = 3$ for $L=4$ results in almost the same FER and BER performance as the optimal values of $S_{\text{Rate-1}} = 3$ and $S_{\text{SPC}} = 4$ as shown in Fig.~\ref{fig:ER1kL4SPC}. As illustrated in Fig.~\ref{fig:ER1kL8SPC} for $L=8$, the selection of $S_{\text{Rate-1}} = 2$ and $S_{\text{SPC}} = 4$ provides similar FER and BER performance as the optimal values of $S_{\text{Rate-1}} = 7$ and $S_{\text{SPC}} = 8$.

\begin{figure}
  \centering
  \hspace{-15pt}
  \input{figures/fer1kL2.tikz}
  \input{figures/ber1kL2.tikz}
  \\
  \hspace{20pt}\ref{perf-legend1kL2}\vspace{2pt}
  \caption{FER and BER performance comparison of Fast-SSCL decoding of $\mathcal{P}(1024,512)$ for $L=2$ and different values of $S_{\text{Rate-1}}$. The CRC length is $16$.}
  \label{fig:ER1kL2}
\end{figure}
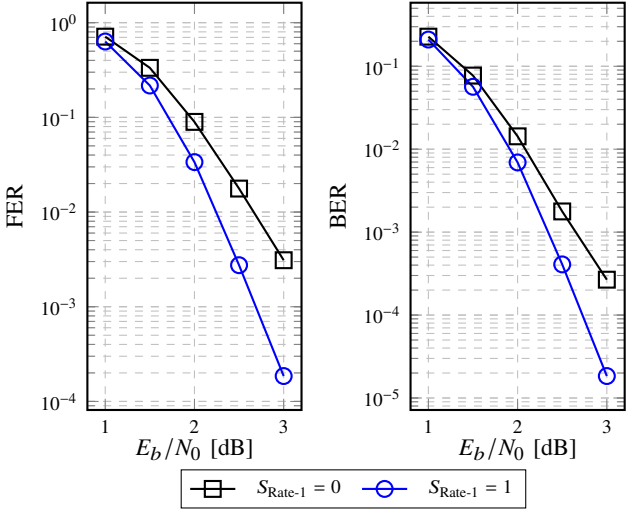

\begin{figure}
  \centering
  \hspace{-15pt}
  \input{figures/fer1kL4.tikz}
  \input{figures/ber1kL4.tikz}
  \\
  \hspace{20pt}\ref{perf-legend1kL4}\vspace{2pt}
  \caption{FER and BER performance comparison of Fast-SSCL decoding of $\mathcal{P}(1024,512)$ for $L=4$ and different values of $S_{\text{Rate-1}}$. The CRC length is $16$.}
  \label{fig:ER1kL4}
\end{figure}

\begin{figure}
  \centering
  \hspace{-15pt}
  \input{figures/fer1kL8.tikz}
  \input{figures/ber1kL8.tikz}
  \\
  \hspace{20pt}\ref{perf-legend1kL8}\vspace{2pt}
  \caption{FER and BER performance comparison of Fast-SSCL decoding of $\mathcal{P}(1024,512)$ for $L=8$ and different values of $S_{\text{Rate-1}}$. The CRC length is $16$.}
  \label{fig:ER1kL8}
\end{figure}

\begin{figure}
  \centering
  \hspace{-15pt}
  \input{figures/fer1kL4SPC.tikz}
  \input{figures/ber1kL4SPC.tikz}
  \\
  \hspace{20pt}\ref{perf-legend1kL4SPC}\vspace{2pt}
  \caption{FER and BER performance comparison of Fast-SSCL-SPC decoding of $\mathcal{P}(1024,512)$ for $L=4$ and different values of $S_\text{Rate-1}$ and $S_\text{SPC}$. The CRC length is $16$.}
  \label{fig:ER1kL4SPC}
\end{figure}
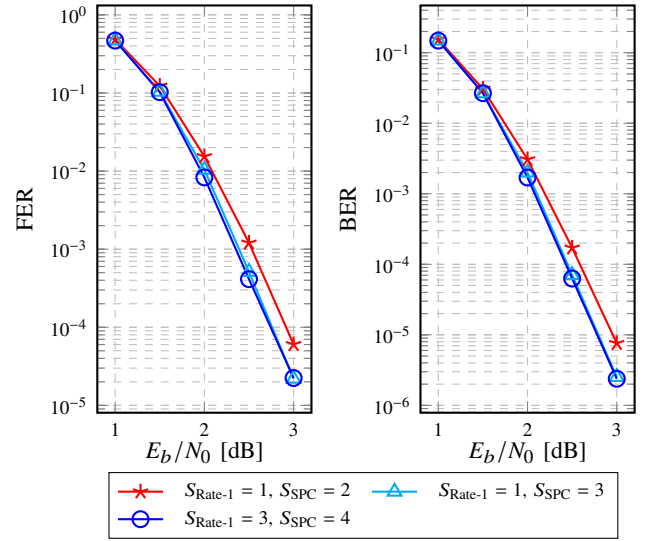

\section{Decoder Architecture}
\label{sec:decarch}

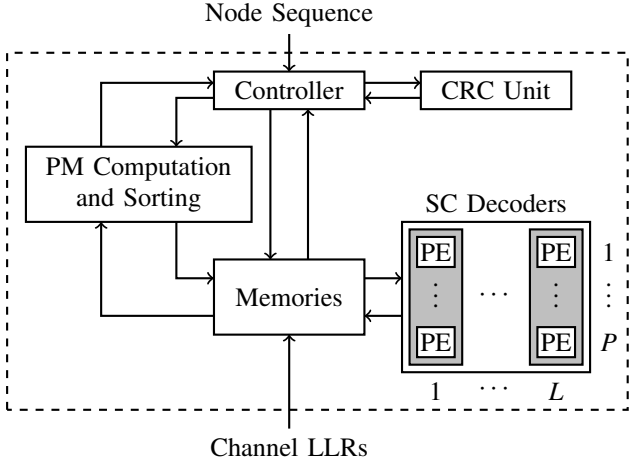
\begin{figure}
  \centering
  \input{figures/decArch.tikz}
  \caption{Decoder architecture.}
  \label{fig:decArch}
\end{figure}

To evaluate the impact of the proposed techniques on a practical case, a SCL-based polar code decoder architecture implementing Fast-SSCL and Fast-SSCL-SPC has been designed. Its basic structure is inspired to the decoders presented in \cite{balatsoukas_SCL_HW,hashemi_SSCL_TCASI}, and it is portrayed in Fig.~\ref{fig:decArch}. The decoding flow follows the one portrayed in Section \ref{sec:prel:SCLDec} for a list size $L$. This means that the majority of the datapath and of the memory are replicated $L$ times, and work concurrently on different candidate codewords and the associated LLR values.

Starting from the tree root, the tree is descended by recursively computing (\ref{eq4}) and (\ref{eq2}) on left and right branches respectively at each tree stage $s$, with a left-first rule. The computations are performed by $L$ sets of $P$ processing elements (PEs), where each set can be considered a standalone SC decoder, and $P$ is a power of $2$. In case $2^s>2P$, (\ref{eq4}) and (\ref{eq2}) require $2^s/(2P)$ time steps to be completed, while otherwise needing a single time step. The updated LLR values are stored in dedicated memories.

The internal structure of PEs is shown in Fig.~\ref{fig:PEArch}. Each PE receives as input two LLR values, outputting one. The computations for both (\ref{eq4}) and (\ref{eq2}) are performed concurrently, and the output is selected according to $i_s$, that represents the $s$-th bit of the index $i$, where $0\le i < N$. The index $i$ is represented with $s_{\max} = \log_2 N$ bits, and identifies the next leaf node to be estimated, and can be composed by observing the path from the root node to the leaf node. From stage $s_{\max}$ down to $0$, for every left branch we set the corresponding bit of $i$ to $0$, and to $1$ for every right branch.

When a leaf node is reached, the controller checks Node Sequence, identifying the leaf node as an information bit or a frozen bit. In case of a frozen bit, the paths are not split, and the bit is estimated only as $0$. All the $L$ path memories are updated with the same bit value, as are the LLR memories and the $\beta$ memories. On the other hand, in case of an information bit, both $0$ and $1$ are considered. The paths are duplicated and the $\PM$s are calculated for the $2L$ candidates according to (\ref{eq7_1}).
They are subsequently filtered through the sorter module, designed for minimum latency. Every $\PM$ is compared to every other in parallel: dedicated control logic uses the resulting signals to return the values of the $\PM$s of the surviving paths and the newly estimated bits they are associated with. The latter are used to update the LLR memories, the $\beta$ memories and the path memories, while also being sent to the CRC calculation module to update the remainder.

All memories in the decoder are implemented as registers: this allows the LLR and $\beta$ values to be read, updated by the PEs, and written back in a single clock cycle. At the same time, the paths are either updated, or split and updated (depending on the constituent code), and the new $\PM$s computed. In the following clock cycle, in case the paths were split, the $\PM$s are sorted, paths are discarded and the CRC value updated. In case paths were not split, the $\PM$s are not sorted, and the CRC update occurs in parallel with the following operation.

\subsection{Memory Structure}
\label{subsec:mem}
\begin{figure}
  \centering
  \input{figures/MemoryArch.tikz}
  \caption{Memory architecture.}
  \label{fig:MemoryArch}
\end{figure}
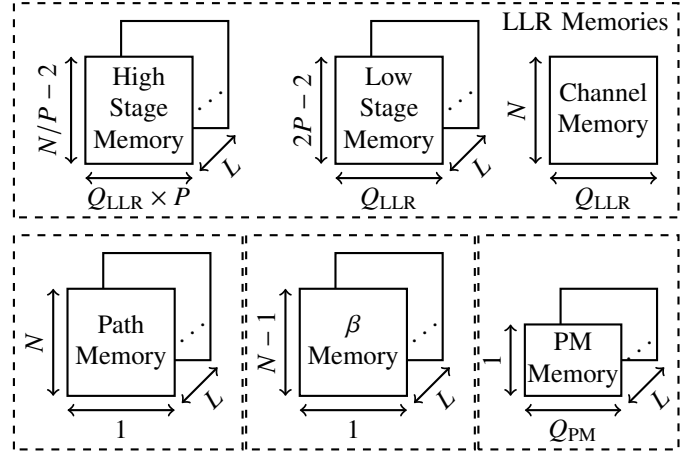

The decoding flow described above relies on a number of memories that are shown in Fig.~\ref{fig:MemoryArch}. The channel memory stores the $N$ LLR values received from the channel at the beginning of the decoding process. Each LLR value is quantized with $Q_{\text{LLR}}$ bits, and represented with sign and magnitude. The high and low stage memories store the intermediate $\alpha$ computed in (\ref{eq4}) and (\ref{eq2}). The high stage memory is used to store LLR values related to stages with nodes of size greater than $P$. The number of PEs determines the number of concurrent (\ref{eq4}) or (\ref{eq2}) that can be performed: for a node in stage $s$, where $2^s>2P$, a total of $2^s/(2P)$ time steps are needed to descend to the lower tree level. The depth of the high stage memory is thus $\sum_{j=\log_2 P+1}^{s_{\max}-1} 2^j/P = N/P-2$, while its width is $Q_{\text{LLR}}\times P$. On the other hand, the low stage memory stores the LLR values for stages where $2^s\le 2P$: the width of this memory is $Q_{\text{LLR}}$, while its depth is defined as $\sum_{j=0}^{\log_2 P - 1} P/2^j = 2P-2$. Both high and low stage memory words are reused by nodes belonging to the same stage $s$, since once a subtree has been completely decoded, its LLR values are not needed anymore. While high and low stage memories are different for each path, the channel LLR values are shared among the $L$ datapaths.
Table~\ref{tab:memAccess} summarizes the memory read and write accesses for the aforementioned LLR memories. When $2^s=2P$, $2P$ LLR values are read from the high stage memory, and the $P$ resulting LLR values are written in the low stage memory. The channel memory is read at $s_{\max}$ only.

\begin{figure}
  \centering
  \input{figures/PEArch.tikz}
  \caption{PE architecture.}
  \label{fig:PEArch}
\end{figure}
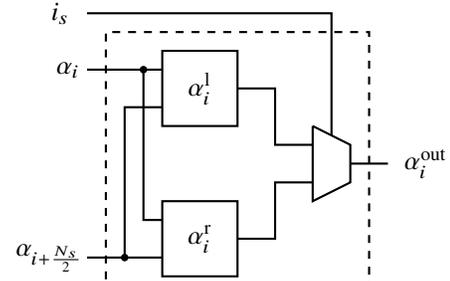

\begin{table}
	\centering
	\caption{LLR Memory Access.}
	\label{tab:memAccess}
		\setlength{\extrarowheight}{1.7pt}
	\begin{tabularx}{.35\textwidth}{ccc}
\toprule

	Stage & READ & WRITE \\

\cmidrule(lr){1-1}
\cmidrule(lr){2-2}
\cmidrule(lr){3-3}

	$s=s_{\max}$ & Channel & High Stage \\
	$\log_2 P + 1 < s < s_{\max}$ & High Stage & High Stage \\
	$s = \log_2 P + 1$ & High Stage & Low Stage \\
	$s < \log_2 P + 1$ & Low Stage & Low Stage \\

\bottomrule
	\end{tabularx}
\end{table}

Each of the $L$ candidate codewords is stored in one of the $N$-bit path memories, updated after every bit estimation. The $\beta$ memories hold the $\beta$ values for each stage from $0$ to $s_{\max}-1$, for a total of $N-1$ bits each. Each time a bit is estimated, all the $\beta$ values it contributes to are concurrently updated. When the decoding of the left half of the SC decoding tree has been completed, the $\beta$ memories are reused for the right half. Finally, the $\PM$ memories store the $L$ $\PM$ values computed in (\ref{eq7_1}).

\subsection{Special Nodes}
\label{subsec:spec}
%Node sequence!
The decoding flow and memory structure described before implement the standard SCL decoding algorithm. The SSCL, SSCL-SPC and the proposed Fast-SSCL and Fast-SSCL-SPC algorithms demand modifications in the datapath to accommodate the simplified computations for \mbox{Rate-0}, \mbox{Rate-1}, \mbox{Rep} and \mbox{SPC} nodes.

As with standard SCL, the pattern of frozen and information bits is known a priori given a polar code structure, the same can be said for special nodes. In the modified architecture, the Node Sequence input in the controller (see Fig.~\ref{fig:decArch}) is not limited to the frozen/information bit pattern, but it includes the type of encountered nodes, their size and the tree stage in which they are encountered. Table~\ref{tab:nodeSequence} summarizes the content of Node Sequence depending on the type of node for SSCL and SSCL-SPC, while in case of Fast-SSCL and Fast-SSCL-SPC Node Sequence is detailed in Table~\ref{tab:nodeSequenceFast}.
The node stage allows the decoder to stop the tree exploration at the right level, and the node type identifies the operations to be performed. Each of the four node types is represented with one or more decoding phases, each of which involves a certain number of codeword bits, identified by the node size parameter. Finally, the frozen bit parameter identifies a bit or set of bits as frozen or not. To limit the decoder complexity, the maximum node stage for special nodes is limited to $s=\log_2 P$, thus the maximum node size is $P$. If the code structure identifies special nodes with node size larger than $P$, they are considered as composed by a set of $P$-size special nodes.

\begin{table}
	\centering
	\caption{Node Sequence Input Information for SSCL and SSCL-SPC.}
	\label{tab:nodeSequence}
		\setlength{\extrarowheight}{1.7pt}
	\begin{tabularx}{.37\textwidth}{lccc}
\toprule

	Node Type & Node Stage & Node Size & Frozen \\

\cmidrule(lr){1-1}
\cmidrule(lr){2-2}
\cmidrule(lr){3-3}
\cmidrule(lr){4-4}

	RATE0 & $s$ &$2^s$& $1$\\
	RATE1 & $s$ &$2^s$& $0$\\
	REP1 & $s$ &$2^s-1$& $1$\\
	REP2 & $s$ &$1$& $0$\\
	DESCEND & Next Node& Next Node& Next Node\\
	LEAF &$0$& $1$ & $0/1$\\

\midrule

	SPC1 & $s$ &$1$& $1$\\
	SPC2 & $s$ &$2^s-1$& $0$\\
	SPC3 & $s$ &$2^s$& $0$\\

\bottomrule
	\end{tabularx}
\end{table}

\begin{table}
	\centering
	\caption{Node Sequence Input Information for Fast-SSCL and Fast-SSCL-SPC.}
	\label{tab:nodeSequenceFast}
		\setlength{\extrarowheight}{1.7pt}
	\begin{tabularx}{.45\textwidth}{lccc}
\toprule

	Node Type & Node Stage & Node Size & Frozen \\

\cmidrule(lr){1-1}
\cmidrule(lr){2-2}
\cmidrule(lr){3-3}
\cmidrule(lr){4-4}

	RATE0 & $s$ &$2^s$& $1$\\
	RATE1-1 & $s$ &$\min(S_\text{Rate-1},2^s)$& $0$\\
	RATE1-2 & $s$ &$2^s-\min(S_\text{Rate-1},2^s)$& $0$\\
	REP1 & $s$ &$2^s-1$& $1$\\
	REP2 & $s$ &$1$& $0$\\
	DESCEND & Next Node& Next Node& Next Node\\
	LEAF &$0$& $1$ & $0/1$\\
\midrule
	SPC1 & $s$ &$1$& $1$\\
	SPC2-1 & $s$ &$\min(S_\text{SPC},2^s)$& $0$\\
	SPC2-2 &$s$ &$2^s-\min(S_\text{SPC},2^s)-1$ &$0$\\
	SPC3 & $s$ &$2^s$& $0$\\

\bottomrule
	\end{tabularx}
\end{table}

\begin{itemize}
\item \mbox{Rate-0} nodes are identified in the Node Sequence with a single decoding phase. No path splitting occurs, and all the $2^s$ node bits are set to $0$. The $\PM$ update requires a single time step, as discussed in \cite{hashemi_SSCL_TCASI}.

\item \mbox{Rate-1} nodes are composed of a single phase in both SSCL and SSCL-SPC, in which paths are split $2^s$ times. In case of Fast-SSCL and Fast-SSCL-SPC, each \mbox{Rate-1} is divided into two phases. The first takes care of the $\min(S_\text{Rate-1},2^s)$ path forks, requiring as many time steps, while the second sets the remaining $2^s-\min(S_\text{Rate-1},2^s)$ bits according to (\ref{eq:hardDecLLR}) and updates the $\PM$ according to (\ref{eq:pmUpdateHW}). This second phase takes a single time step.

\item \mbox{Rep} nodes are identified by two phases in the Node Sequence, the first of which takes care of the $2^s-1$ frozen bits similarly as \mbox{Rate-0} nodes do, and the second estimates the single information bit. Each of these two phases lasts a single time step.

\item \mbox{SPC} nodes are split in three phases in the original SSCL-SPC formulation. The first phase takes care of the frozen bit, and computes both (\ref{eq:SPCmin}) and (\ref{eq:SPCgamma}), initializing the $\PM$ as (\ref{eq:SPCPM0}) in a time step. The extraction of the least reliable bit in (\ref{eq:SPCmin}) is performed through a comparison tree that carries over both the index and the value of the LLR.

The second phase estimates the $2^s-1$ information bits, splitting the path as many times in as many time steps. During this phase, each time a bit is estimated, it is XORed with the previous $\beta$ values: this operation is useful to compute (\ref{eq:SPCbitCorrection}). The update of $\beta_{i_{\min}}$ is finally performed in the third phase, that takes a single time step. Moving to Fast-SSCL-SPC, the second \mbox{SPC} phase is split in two, similarly to what happens to the \mbox{Rate-1} node.

\item \mbox{Descend} is a non-existing node type that is inserted for one clock cycle in Node Sequence for control purposes after every special node. The node size and stage associated with this label are those of the following node. The \mbox{Descend} node type is used by the controller module.

\item \mbox{Leaf} nodes identify all nodes that can be found at $s=0$, for which the standard SCL algorithm applies. 

\end{itemize}

The decoding of special nodes requires a few major changes in the decoder architecture.
\begin{itemize}
 \item \emph{Path Memory}: each path memory is an array of $N$ registers, granting concurrent access to all bits with a $1$-bit granularity. In SCL, the path update is based on the combination of a write enable signal, the codeword bit index $i$ that acts as a memory address, and the value of the estimated bit after the $\PM$s have been sorted and the surviving paths identified. 
Fig.~\ref{fig:pathMemMod} shows the path memory access architecture for Fast-SSCL-SPC. Unlike SCL, the path memory is not always updated with the estimated bit $\hat{u}$. Thus, the SCL datapath is bypassed according to the node type. When Node Sequence identifies RATE0, REP1 and SPC1 nodes that consider frozen bits, the path memory is updated with $0$ values. The estimated bit $\hat{u}$ is chosen as input for RATE1-1, REP2, SPC2-2 and LEAF nodes, where the path is split. RATE1-2 and SPC2-2 nodes estimate the bits through hard decision on the LLR values, while in the SPC3 case the update considers the result of (\ref{eq:SPCbitCorrection}).
At the same time. whenever the estimated bits are more than one, the corresponding bits in the path memory must be concurrently updated. Thus, the address becomes a range of addresses for RATE0, RATE1-2, REP1 and SPC2-2.
 
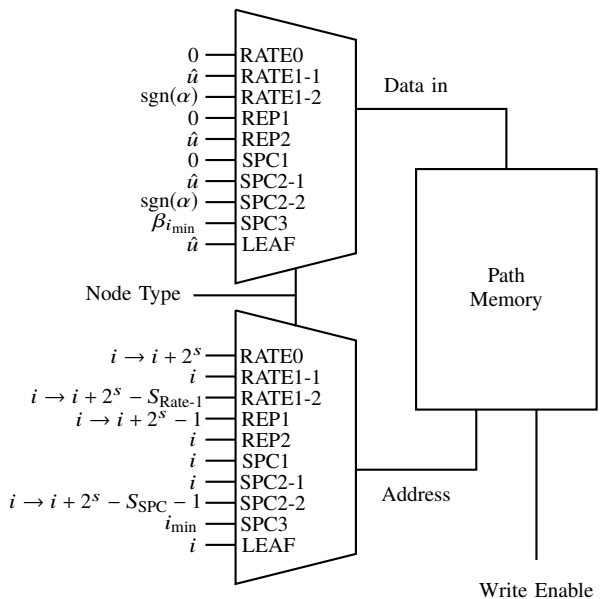
\begin{figure}
  \centering
  \input{figures/PathMemMod.tikz}
  \caption{Path memory access architecture for Fast-SSCL-SPC.}
  \label{fig:pathMemMod}
\end{figure}
 
 \item \emph{$\beta$ Memory}: the update of this memory depends on the value of the estimated bit. In order to limit the latency cost of these computations, concurrently to the estimation of $\hat{u}$, the updated values of all the bits of the $\beta$ memory are computed assuming both $\hat{u}=0$ and $\hat{u}=1$. The actual value of $\hat{u}$ is used as a selection signal to decide on the two alternatives. The $\beta$ memory in SCL, unlike the path memory, already foresees the concurrent update of multiple entries that are selected based on the bit index $i$. Given an estimated leaf node, the $\beta$ values of all the stages that it affects are updated: in fact, since as shown in (\ref{eq3}) the update of $\beta$ values is at most a series of XORs, it is possible to distribute this operation in time. The same can be said of multi-bit (\ref{eq3}) updates.
 To implement Fast-SSCL-SPC, the $\beta$ update selection logic must be modified to foresee the special nodes, similar to that portrayed in Fig.~\ref{fig:pathMemMod} for the path memory. For RATE0, REP1, and SPC1, the $\hat{u}=0$ update is always selected. RATE1-1, REP2, SPC2-1 and LEAF nodes maintain the standard SCL selection based on the actual value of $\hat{u}$. The update for SPC3 case is based on $\beta_{i_{\min}}$. For RATE1-2 and SPC1-2, the selection is based on the XORed sign bits of the LLR values read from the memory.

 \item \emph{$\PM$ Calculation}: this operation is performed, in the original SCL architecture and for leaf nodes in general according to (\ref{eq7_1}). The paths and associated $\PM$s are split and sorted every time an information bit is estimated, while $\PM$s are updated without sorting when frozen bits are encountered. While the sorting architecture remains the same, the implementation of the proposed algorithm requires a different $\PM$ update structure for each special node. Unlike with leaf nodes, the LLR values needed for the $\PM$ update in special nodes are not the output of PEs, and are read directly from the LLR memories. Additional bypass logic is thus needed. For RATE0 and REP1, (\ref{eq:Rate0:HWF}) and (\ref{eq:Rep:HWF}) require a summation over up to $P$ values, while SPC1 nodes need to perform the minimum $\alpha$ search (\ref{eq:SPCmin}): these operations are tackled through adder and comparator trees. RATE1-1, REP2 and SPC2-1 $\PM$ updates are handled similarly to the leaf node case, since a single bit at a time is being estimated. RATE1-2, SPC2-2 and SPC3 do not require any $\PM$ to be updated. 

 \item \emph{CRC Calculation}: the standard SCL architecture foresees the estimation of a single bit at a time. Thus, the CRC is computed sequentially. However, \mbox{Rate-0} and \mbox{Rep} nodes in SSCL and SSCL-SPC estimate up to $P$ and $P-1$ bits concurrently. Thus, for the CRC operation not to become a latency bottleneck, the CRC calculation must be parallelized by updating the remainder. Following the idea presented in \cite{Condo_CRC}, it is possible to allow for variable input sizes with a high degree of resource sharing and limited overall complexity. The circuit is further simplified by the fact that both \mbox{Rate-0} and \mbox{Rep} nodes guarantee that the estimated bit values are all $0$.
Fig.~\ref{fig:CRCArch} shows the modified CRC calculation module in case $P=64$, where $N_{\text{CRC}}$ represents the number of concurrently estimated bits: the estimated bit can be different from $0$ only in case of leaf nodes and $s=1$ \mbox{Rep} nodes, for which a single bit is estimated in any case.

The Fast-SSCL and Fast-SSCL-SPC architectures follow the same idea, but require additional logic. RATE1-2 and SPC2-2 nodes introduce new degrees of parallelism, as up to $P-S_\text{Rate-1}$ and $P-S_\text{SPC}$ bits are updated at the same time. Moreover, it is not possible to assume that these bits are $0$ as with RATE0 and REP1. The value of the estimated bit must be taken into account, leading to increased complexity.

\begin{figure}
  \centering
  \input{figures/CRCArch.tikz}
  \caption{CRC architecture for SSCL and SSCL-SPC.}
  \label{fig:CRCArch}
\end{figure}
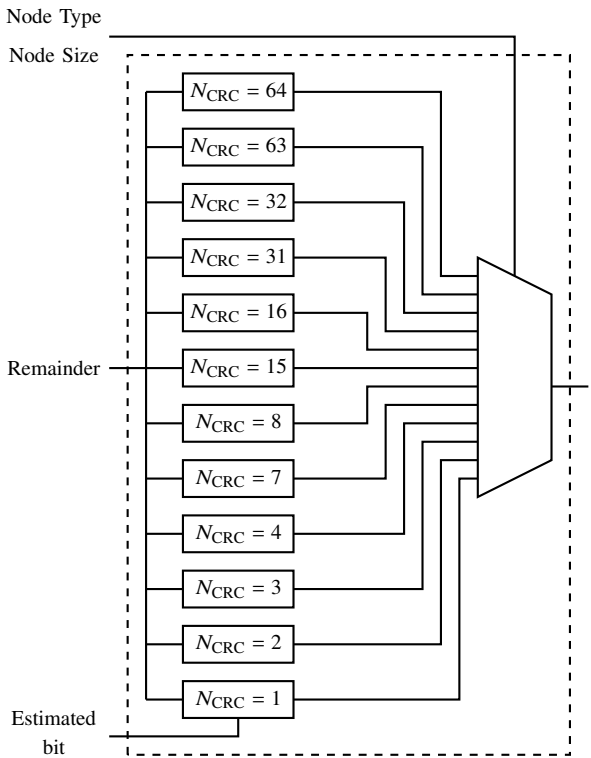

 \item \emph{Controller}: this module in the SCL architecture is tasked with the generation of memory write enables, the update of the codeword bit index $i$ and the stage tracker $s$, along with the LLR memory selection signals according to Table~\ref{tab:memAccess} and path enable and duplication signals. It implements a finite state machine that identifies the status of the decoding process.
 The introduction of special nodes demands that most of the control signal generation logic is modified. Of particular importance is the fact that, in the SCL architecture, the update of $i$ is bound to having reached a leaf node, i.e. $s=0$. In Fast-SSCL-SPC, it is instead linked to $s$ being equal to the special node stage. The index $i$ is moreover incremented of the amount of bits estimated in a single time step, depending on the type of node. Memory write enables are also bound to having reached the special node stage, and not only to $s=0$.
\end{itemize}

\section{Results}
\label{sec:results}

\subsection{Hardware Implementation}

\begin{table*}
	\centering
	\caption{TSMC 65~nm Implementation Results for $\mathcal{P}(1024,512)$ and $P=64$.}
	\label{tab:implementations}
		\setlength{\extrarowheight}{1.7pt}
%\scriptsize
	\begin{tabularx}{.6\textwidth}{ccccYYY}
\toprule

Implementation & $L$ & $S_{\text{Rate-1}}$ & $S_{\text{SPC}}$ & \begin{tabular}{c}Area\\{[mm$^2$]}\end{tabular} & \begin{tabular}{c}Frequency\\{[MHz]}\end{tabular} & \begin{tabular}{c}Throughput\\{[Mb/s]}\end{tabular} \\

\cmidrule(lr){1-1}
\cmidrule(lr){2-2}
\cmidrule(lr){3-4}
\cmidrule(lr){5-7}

\multirow{3}{*}{\bf SCL} & $2$ & \ding{55} & \ding{55} & $0.599$ & $1031$ & $389$ \\

 & $4$ & \ding{55} & \ding{55} & $0.998$ & $961$& $363$ \\

 & $8$ & \ding{55} & \ding{55} & $2.686$& $722$& $272$\\

\midrule

\multirow{3}{*}{\bf SSCL} & $2$ & \ding{55} & \ding{55} & $0.643$ & $1031$ & $1108$ \\

 & $4$ & \ding{55} & \ding{55} & $1.192$ & $961$ & $1033$ \\
 
 & $8$ & \ding{55} & \ding{55} & $2.958$ & $722$ & $776$ \\

\midrule

\multirow{3}{*}{\bf SSCL-SPC} & $2$ & \ding{55} & \ding{55} & $0.684$ & $1031$ & $1229$ \\

 & $4$ & \ding{55} & \ding{55} & $1.223$  &$961$ &$1146$\\

 & $8$ & \ding{55} & \ding{55} &$3.110$ & $722$ & $861$\\

\midrule

\multirow{5}{*}{\bf Fast-SSCL} & $2$ & $1$ & \ding{55} &$0.871$ &$885$ & $1579$\\

 & $4$ & $1$ & \ding{55} &$1.536$ & $840$ & $1499$\\
 & $4$ & $3$ & \ding{55} &$1.511$ & $840$& $1446$\\

 & $8$ & $2$ & \ding{55} & $3.622$& $722$ & $1053$\\
 & $8$ & $7$ & \ding{55} & $3.588$& $722$ & $827$\\

\midrule

\multirow{5}{*}{\bf Fast-SSCL-SPC} & $2$ & $1$ & $2$ & $1.048$&$885$ & $1861$\\

 & $4$ & $1$ & $3$ & $1.822$ &$840$ & $1608$\\
 & $4$ & $3$ & $4$ & $1.797$ &$840$ & $1338$\\

 & $8$ & $2$ & $4$ & $3.975$ &$722$ & $1198$\\
 & $8$ & $7$ & $8$ & $3.902$ &$722$ & $959$\\

\bottomrule
	\end{tabularx}
\end{table*}

The architecture designed in Section~\ref{sec:decarch} has been described in the VHDL language and synthesized in TSMC 65 nm CMOS technology. Implementation results are provided in Table~\ref{tab:implementations} for different decoders: along with the Fast-SSCL and Fast-SSCL-SPC described in this work, the SCL, SSCL and SSCL-SPC decoders proposed in \cite{hashemi_SSCL_TCASI} are presented as well. Each decoder has been synthesized with three list sizes ($L=2, 4, 8$), while the Fast-SSCL and Fast-SSCL-SPC architectures have been synthesized for considering different combinations of $S_{\text{Rate-1}}$ and $S_{\text{SPC}}$, as portrayed in Section~\ref{sec:FSSCL:speed}. Quantization values are the same used in \cite{hashemi_SSCL_TCASI}, i.e. $6$ bits for LLR values and $8$ bits for $\PM$s, with two fractional bits each. All memory elements have been implemented through registers and the area results include both net area and cell area. The reported throughput is coded.

All Fast-SSCL and Fast-SSCL-SPC, regardless of the value of $S_{\text{Rate-1}}$ and $S_{\text{SPC}}$, show a substantial increase in area occupation with respect to SSCL and SSCL-SPC. The main contributing factors to the additional area overhead are three:
\begin{itemize}
 \item In SSCL and SSCL-SPC, the CRC computation needs to be parallelized, since in \mbox{Rep} and \mbox{Rate-0} nodes multiple bits are updated at the same time. However, the bit value is known at design time, since they are frozen bits. This, along with the fact that $0$ is neutral in the XOR operations required by CRC calculation, limits the required additional area overhead. On the contrary, in Fast-SSCL and Fast-SSCL-SPC, \mbox{Rate-1} and \mbox{SPC} nodes update multiple bits within the same time step (SPC2-2 and RATE1-2 stages). In these cases, however, they are information bits, whose values cannot be known at design time: the resulting parallel CRC tree is substantially wider and deeper than the ones for \mbox{Rate-0} and \mbox{Rep} nodes. Moreover, with increasing number of CRC trees, the selection logic becomes more cumbersome.
 \item A similar situation is encountered for the $\beta$ memory update signal. As described in the previous section, the $\beta$ memory update values are computed assuming both estimated values, and the actual value of $\hat{u}$ is used as a selection signal. In SSCL and SSCL-SPC the multiple-bit update does not constitute a problem since all the estimated bits are $0$ and the $\beta$ memory content does not need to be changed. On the contrary, in Fast-SSCL and Fast-SSCL-SPC, the value of the estimated information bits might change the content of the $\beta$ memory. Moreover, since $\beta$ is computed as (\ref{eq3}), the update of $\beta$ bits depends on previous bits as well as the newly estimated ones. Thus, an XOR tree is necessary to compute the right selection signal for every information bit estimated in SPC2-2 and RATE1-2 stages.
 \item The aforementioned modifications considerably lengthen the system critical path. In case of large code length, small list size, or large $P$, the critical path starts in the controller module, in particular in the high stage memory addressing logic, goes through the multiplexing structure that routes LLR values to the PEs, and ends after the $\PM$ update. In case of large list sizes or short code length, the critical path passes through the PM sorting and path selection logic, and through the parallel CRC computation. Thus, pipeline registers have been inserted to lower the impact of critical path, at the cost of additional area occupation.
\end{itemize}

Fast-SSCL and Fast-SSCL-SPC implementations show consistent throughput improvements with respect to previously proposed architectures. The gain is lower than what is shown to be theoretically achievable in Fig.~\ref{figTimeReq}. This is due to the aforementioned pipeline stages, that increase the number of steps needed to complete the decoding of component codes.

\subsection{Comparison with Previous Works}
\label{subsec:compSoA}

The Fast-SSCL-SPC hardware implementation presented in this paper for $\mathcal{P}(1024,512)$ and $P=64$ is compared with the state-of-the-art architectures in \cite{yuan_multibit_LLR,xiong_symbol,lin_SCL,xiong_multimode,hashemi_SSCL_TCASI} and the results are provided in Table~\ref{tab:compare}. The architectures presented in \cite{xiong_symbol,lin_SCL,xiong_multimode} were synthesized based on 90~nm technology: for a fair comparison, their results have been converted to 65~nm technology using a factor of $90/65$ for the frequency and a factor of $\left(65/90\right)^2$ for the area. The synthesis results in \cite{yuan_multibit_LLR} were carried out in 65~nm technology but reported in 90~nm technology. Therefore, a reverse conversion was applied to convert the results back to 65~nm technology.

\begin{table*}
	\centering
	\caption{Comparison with State-of-the-Art Decoders.}
	\label{tab:compare}
		\setlength{\extrarowheight}{1.7pt}
%\scriptsize
	\begin{tabularx}{.9\textwidth}{cYYYYYYYYYYY}
\toprule
 & \multicolumn{3}{c}{This work} & \cite{yuan_multibit_LLR} & \cite{xiong_symbol}\textsuperscript{\textdagger} & \multicolumn{2}{c}{\cite{lin_SCL}\textsuperscript{\textdagger}} & \cite{xiong_multimode}\textsuperscript{\textdagger} & \multicolumn{3}{c}{\cite{hashemi_SSCL_TCASI}} \\

\cmidrule(lr){1-1}
\cmidrule(lr){2-4}
\cmidrule(lr){5-5}
\cmidrule(lr){6-6}
\cmidrule(lr){7-8}
\cmidrule(lr){9-9}
\cmidrule(lr){10-12}

$L$ & $2$ & $4$ & $8$ & $4$ & $4$ & $2$ & $4$ & $4$ & $2$ & $4$ & $8$ \\

$P$ & $64$ & $64$ & $64$ & $64$ & $64$ & $64$ & $64$ & $256$ & $64$ & $64$ & $64$ \\

\midrule

Area [mm$^2$] &$1.048$ &$1.822$ &$3.975$ & $0.62$ & $0.73$ & $1.03$ & $2.00$ & $0.99$ & $0.68$ & $1.22$ & $3.11$ \\

Frequency [MHz] &$885$ &$840$ &$722$ & $498$ & $692$ & $586$ & $558$ & $566$ & $1031$ & $961$ & $722$ \\

Throughput [Mb/s] &$1861$ &$1608$ &$1198$ & $935$ & $551$ & $1844$ & $1578$ & $1515$ & $1229$ & $1146$ & $861$ \\

Latency [$\upmu$s] & $0.55$& $0.64$ & $0.85$& $1.10$ & $1.86$ & $0.57$ & $0.66$ & $0.69$ & $0.83$ & $0.89$ & $1.19$ \\

Area Efficiency [Mb/s/mm$^2$] & $1776$& $883$& $301$& $1508$ & $755$ & $1790$ & $789$ & $1530$ & $1807$ & $939$ & $277$ \\
\midrule[\heavyrulewidth]
\multicolumn{10}{l}{\textsuperscript{\textdagger}\footnotesize{The results are originally based on TSMC 90~nm technology and are scaled to TSMC 65~nm technology.}} \\
\bottomrule
	\end{tabularx}
\end{table*}

The architecture in this paper shows $72\%$ higher throughput and $42\%$ lower latency with respect to the multibit decision SCL decoder architecture of \cite{yuan_multibit_LLR} for $L=4$. However, the area occupation of \cite{yuan_multibit_LLR} is smaller, leading to a higher area efficiency than the design in this paper.

The symbol-decision SCL decoder architecture of \cite{xiong_symbol} shows lower area occupation than the design in this paper for $L=4$ but it comes at the cost of lower throughput and higher latency. Our decoder architecture achieves $192\%$ higher throughput and $66\%$ lower latency than \cite{xiong_symbol} which resulted in $17\%$ higher area efficiency.

The high throughput SCL decoder architecture of \cite{lin_SCL} for $L=2$ requires lower area occupation than our design but it comes at the expense of lower throughput and higher latency. Moreover, the design in \cite{lin_SCL} relies on parameters that need to be tuned for each code, and it is shown in \cite{lin_SCL} that a change of code can result in more than $0.2$~dB error-correction performance loss.
For $L=4$, our decoder not only achieves higher throughput and lower latency than \cite{lin_SCL}, but also it occupies a smaller area. This in turn yields a $12\%$ increase in the area efficiency in comparison with \cite{lin_SCL}.

The multimode SCL decoder in \cite{xiong_multimode} relies on a higher number of PEs than our design: nevertheless, it yields lower throughput and higher latency than the architecture proposed in this paper for $L=4$. It should be noted that \cite{xiong_multimode} is based on the design presented in \cite{lin_SCL}, whose code-specific parameters may lead to substantial error-correction performance degradation. On the contrary, the design in this paper is targeted for speed and flexibility and can be used to decode any polar code of any length.

Compared to our previous work \cite{hashemi_SSCL_TCASI}, that has the same degree of flexibility of the proposed design, this decoder achieves $51\%$ higher throughput and $34\%$ lower latency for $L=2$, and $40\%$ higher throughput and $28\%$ lower latency for $L=4$. However, the higher area occupation of the new design yields lower area efficiencies than \cite{hashemi_SSCL_TCASI} for $L=\{2,4\}$. For $L=8$, the proposed design has $39\%$ higher throughput and $29\%$ lower latency than \cite{hashemi_SSCL_TCASI}, which results in $9\%$ increase in area efficiency. The reason is that for $L=8$, the sorter is quite large and falls on the critical path. Consequently, the maximum achievable frequency for the proposed design is limited by the sorter and not by \mbox{Rate-1} and \mbox{SPC} nodes as opposed to the $L=\{2,4\}$ case. This results in the same maximum achievable frequency for both designs, hence, higher throughput and area efficiency.

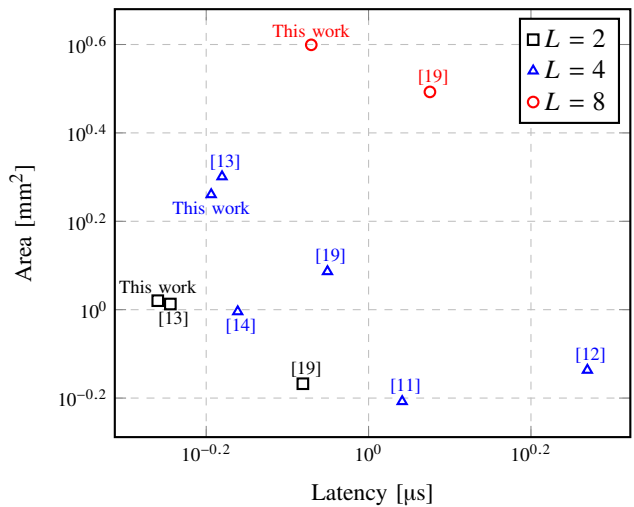
\begin{figure}
  \centering
  \input{figures/AreaLatencyL2.tikz}
  \caption{Comparison with state-of-the-art decoders.}
  \label{fig:ALL2}
\end{figure}

Fig.~\ref{fig:ALL2} plots the area occupation against the decoding latency for all the decoders considered in Table~\ref{tab:compare}. For each value of $L$, the design proposed in this work have the shortest latency, shown by their leftmost position on the graph.

\section{Conclusion} \label{sec:conc}

In this work, we have proven that the list size in polar decoders sets a limit to the useful number of path forks in \mbox{Rate-1} and \mbox{SPC} nodes. We thus propose Fast-SSCL and Fast-SSCL-SPC polar code decoding algorithms that, depending on $L$ and the number of performed path forks, can reduce the number of required time steps of more than $75\%$ at no error-correction performance cost. Hardware architectures for the proposed algorithms have been described and implemented in CMOS 65~nm technology. They have a very high degree of flexibility and can decode any polar code, regardless of its rate. The proposed decoder is the fastest SCL-based decoder in literature: sized for $N=1024$ and $L=2$, it yields a $1.861$ Gb/s throughput with an area occupation of $1.048$~mm$^2$. The same design, sized for $L=4$ and $L=8$, leads to throughputs of $1.608$ Gb/s and $1.198$ Gb/s, and areas of $1.822$~mm$^2$ and $3.975$~mm$^2$, respectively.

\appendices

\section{Proof of Theorem \ref{th:maxEstimateSPC}} \label{sec:app2}

\begin{proof}
In order to prove Theorem \ref{th:maxEstimateSPC}, we note that the first step is to initialize the $\PM$s based on (\ref{eq:SPCPM0}). Therefore, the least reliable bit needs to be estimated first. For the bits other than the least reliable bit, the $\PM$s are updated based on (\ref{eq:SPCPM}). However, the term $(1-2\gamma)|\alpha_{i_{\min}}|$ is constant for all the bit estimations in the same path. Therefore, we can define a new set of $N_s-1$ LLR values as
\begin{equation}
\alpha_{i_m} = \alpha_i + \sgn(\alpha_i)(1-2\gamma)|\alpha_{i_{\min}}| \text{,}
\end{equation}
for $i \neq i_{\min}$ and $0 \leq i_m < N_s-1$, which results in 
\begin{equation}
|\alpha_{i_m}| = |\alpha_i| + (1-2\gamma)|\alpha_{i_{\min}}| \text{.}
\end{equation}
The problem is now reduced to a \mbox{Rate-1} node of length $N_s-1$ which, with the result of Theorem \ref{th:maxEstimate}, can be decoded by considering only $\min(L-1,N_s-1)$ path splitting. Adding the bit estimation for $i_{\min}$, SPC nodes can be decoded by splitting paths $\min(L,N_s)$ times while guaranteeing the same results as in \mbox{SSCL-SPC}. Theorem \ref{th:maxEstimateSPC} is consequently proven.
\end{proof}

\begin{IEEEbiography}[{\includegraphics[width=1in,height=1.25in,clip,keepaspectratio]{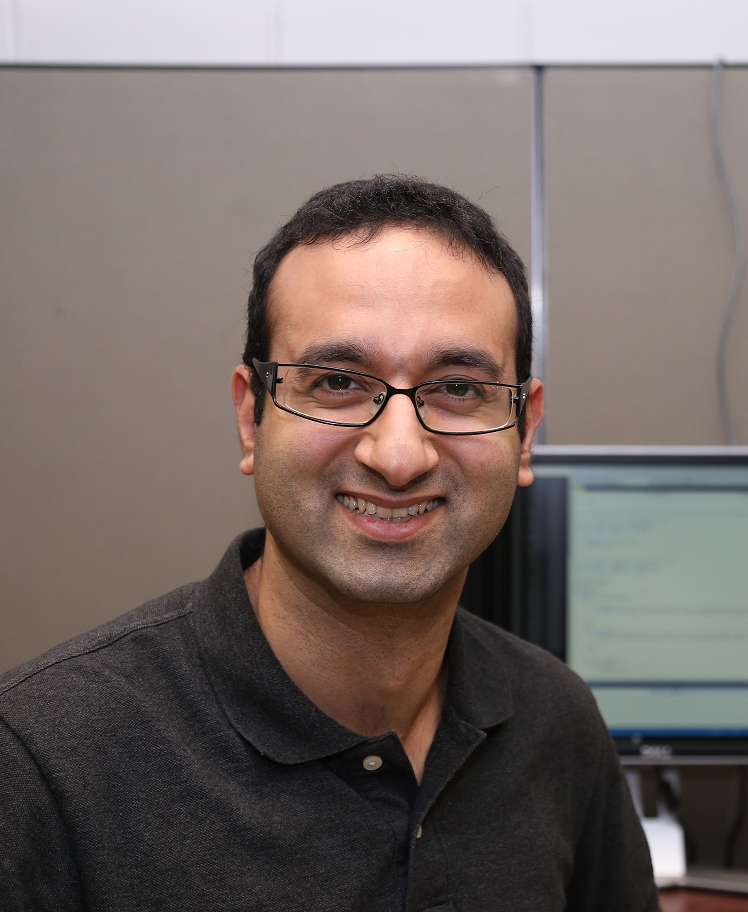}}]{Seyyed Ali Hashemi}
was born in Qaemshahr, Iran. He received the B.Sc. degree in electrical engineering from Sharif University of Technology, Tehran, Iran, in 2009 and the M.Sc. degree in electrical and computer engineering from the University of Alberta, Edmonton, AB, Canada, in 2011. He is currently working toward the Ph.D. degree in electrical and computer engineering at McGill University, Montr\'eal, QC, Canada. He was the recipient of a Best Student Paper Award at the 2016 IEEE International Symposium on Circuits and Systems (ISCAS 2016). His research interests include error-correcting codes, hardware architecture optimization, and VLSI implementation of digital signal processing systems.
\end{IEEEbiography}

\begin{IEEEbiography}[{\includegraphics[width=1in,height=1.25in,clip,keepaspectratio]{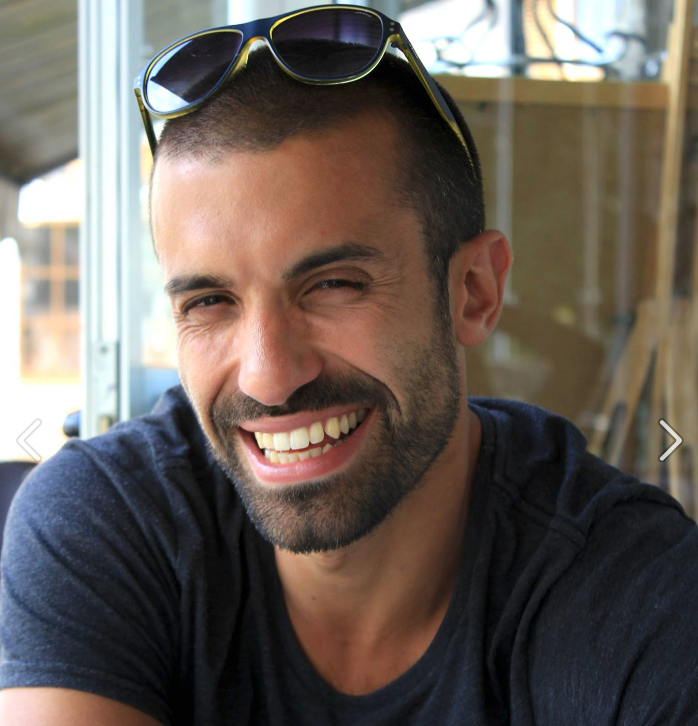}}]{Carlo Condo}
received the M.Sc. degree in electrical and computer engineering from Politecnico di Torino, Italy, and the University of Illinois at Chicago, IL, USA, in 2010. He received the Ph.D. degree in electronics and telecommunications engineering from Politecnico di Torino and Telecom Bretagne, France, in 2014. Since 2015, he has been a postdoctoral fellow at the ISIP Laboratory, McGill University. His Ph.D. thesis was awarded a mention of merit as one of the five best of 2013/2014 by the GE association, and he has been the recipient of two conference best paper awards (SPACOMM 2013 and ISCAS 2016). His research is focused on channel coding, design and implementation of encoder and decoder architectures, and digital signal processing.
\end{IEEEbiography}

\begin{IEEEbiography}[{\includegraphics[width=1in,height=1.25in,clip,keepaspectratio]{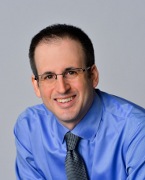}}]{Warren J. Gross}
(SM'10) received the B.A.Sc. degree in electrical engineering from the University of Waterloo, Waterloo, ON, Canada, in 1996, and the M.A.Sc. and Ph.D. degrees from the University of Toronto, Toronto, ON, Canada, in 1999 and 2003, respectively. Currently, he is Professor and Associate Chair (Academic Affairs) with the Department of Electrical and Computer Engineering, McGill University, Montr\'eal, QC, Canada. His research interests are in the design and implementation of signal processing systems and custom computer architectures.

Dr. Gross served as Chair of the IEEE Signal Processing Society Technical Committee on Design and Implementation of Signal Processing Systems. He has served as General Co-Chair of IEEE GlobalSIP 2017 and IEEE SiPS 2017, and as Technical Program Co-Chair of SiPS 2012. He has also served as organizer for the Workshop on Polar Coding in Wireless Communications at WCNC 2017, the Symposium on Data Flow Algorithms and Architecture for Signal Processing Systems (GlobalSIP 2014) and the IEEE ICC 2012 Workshop on Emerging Data Storage Technologies. Dr. Gross served as Associate Editor for the IEEE Transactions on Signal Processing and currently is a Senior Area Editor. Dr. Gross is a Senior Member of the IEEE and a licensed Professional Engineer in the Province of Ontario.
\end{IEEEbiography}

\end{document}

%% file: figures/sc-dec.tikz
\begin{tikzpicture}[scale=2, thick]
\newcommand\Triangle[1]{-- ++(0:2*#1) -- ++(120:2*#1) --cycle}
\newcommand\Square[1]{+(-#1,-#1) rectangle +(#1,#1)}

  \draw (0,0) circle [radius=.05];
  
  \draw (-.05,0) -- (.05,0);
  \draw (0,-.05) -- (0,.05);

  \draw (-1.05,-.55) \Triangle{.05};
  \draw (1,-.5) \Square{.05};
%  \draw [gray, very thick] (-1,-.5) circle [radius=.05];
%  \fill [gray, very thick] (1,-.5) circle [radius=.05];

  \draw (-1.5,-1) circle [radius=.05];
  \draw (-.55,-1.05) \Triangle{.05};
  \draw (.5,-1) \Square{.05};
%  \draw [gray, very thick] (-.5,-1) circle [radius=.05];
%  \fill [gray, very thick] (.5,-1) circle [radius=.05];
  \fill (1.5,-1) circle [radius=.05];

  \draw (-1.75,-1.5) circle [radius=.05];
  \draw (-1.25,-1.5) circle [radius=.05];
  \draw (-.75,-1.5) circle [radius=.05];
  \fill (-.25,-1.5) circle [radius=.05];
  \draw (.25,-1.5) circle [radius=.05];
  \fill (.75,-1.5) circle [radius=.05];
  \fill (1.25,-1.5) circle [radius=.05];
  \fill (1.75,-1.5) circle [radius=.05];

  \node at (-1.75,-1.7) {$\hat{u}_0$};
  \node at (-1.25,-1.7) {$\hat{u}_1$};
  \node at (-.75,-1.7) {$\hat{u}_2$};
  \node at (-.25,-1.7) {$\hat{u}_3$};
  \node at (.25,-1.7) {$\hat{u}_4$};
  \node at (.75,-1.7) {$\hat{u}_5$};
  \node at (1.25,-1.7) {$\hat{u}_6$};
  \node at (1.75,-1.7) {$\hat{u}_7$};

  \draw (0,-.05) -- (-1,-.45);
  \draw (0,-.05) -- (1,-.45);

  \draw (-1,-.55) -- (-1.5,-.95);
  \draw (-1,-.55) -- (-.5,-.95);
  \draw (1,-.55) -- (.5,-.95);
  \draw (1,-.55) -- (1.5,-.95);

  \draw (-1.5,-1.05) -- (-1.75,-1.45);
  \draw (-1.5,-1.05) -- (-1.25,-1.45);
  \draw (-.5,-1.05) -- (-.75,-1.45);
  \draw (-.5,-1.05) -- (-.25,-1.45);
  \draw (.5,-1.05) -- (.25,-1.45);
  \draw (.5,-1.05) -- (.75,-1.45);
  \draw (1.5,-1.05) -- (1.25,-1.45);
  \draw (1.5,-1.05) -- (1.75,-1.45);

  \draw [very thin,gray,dashed] (-2,0) -- (2,0);
  \draw [very thin,gray,dashed] (-2,-.5) -- (2,-.5);
  \draw [very thin,gray,dashed] (-2,-1) -- (2,-1);
  \draw [very thin,gray,dashed] (-2,-1.5) -- (2,-1.5);

  \draw [->] (-.12,-.05) -- (-1,-.4) node [above=-.1cm,midway,rotate=25] {$\bm{\alpha}$};
  \draw [->] (-.88,-.45) -- (0,-.1) node [below=-.1cm,midway,rotate=25] {$\bm{\beta}$};

  \draw [->] (-1.06,-.55) -- (-1.5,-.9) node [above=-.1cm,midway,rotate=40] {$\bm{\alpha}^{\text{l}}$};
  \draw [->] (-1.44,-.95) -- (-1.0,-0.6) node [below=-.1cm,midway,rotate=40] {$\bm{\beta}^{\text{l}}$};

  \draw [<-] (-.94,-.55) -- (-.5,-.9) node [above=-.1cm,midway,rotate=-40] {$\bm{\beta}^{\text{r}}$};
  \draw [<-] (-.56,-.95) -- (-0.975,-.625) node [below=-.1cm,midway,rotate=-40] {$\bm{\alpha}^{\text{r}}$};

\end{tikzpicture}

%% file: figures/sscl-dec.tikz
\begin{tikzpicture}[scale=1, thick]
\newcommand\Triangle[1]{-- ++(0:2*#1) -- ++(120:2*#1) --cycle}
\newcommand\Square[1]{+(-#1,-#1) rectangle +(#1,#1)}

  \draw (0,0) circle [radius=.1];
  
  \draw (-.1,0) -- (.1,0);
  \draw (0,-.1) -- (0,.1);

  \draw (-1.1,-.6) \Triangle{.1};
  \draw (1,-.5) \Square{.1};

  \draw (0,-.1) -- (-1,-.4);
  \draw (0,-.1) -- (1,-.4);
  
  \draw (.4,-1.1) \Triangle{.1};
  \fill (1.5,-1) circle [radius=.1];
  
  \draw (1,-.6) -- (.5,-.9);
  \draw (1,-.6) -- (1.5,-.9);

  \draw [very thin,gray,dashed] (-2,0) -- (2,0);
  \draw [very thin,gray,dashed] (-2,-.5) -- (2,-.5);
  \draw [very thin,gray,dashed] (-2,-1) -- (2,-1);

  \node at (-1,-.8) {Rep};
  \node at (.5,-1.3) {Rep};
  \node at (1.5,-1.3) {Rate-1};

\end{tikzpicture}

%% file: figures/ssclspc-dec.tikz
\begin{tikzpicture}[scale=1, thick]
\newcommand\Triangle[1]{-- ++(0:2*#1) -- ++(120:2*#1) --cycle}
\newcommand\Square[1]{+(-#1,-#1) rectangle +(#1,#1)}

  \draw (0,0) circle [radius=.1];
  
  \draw (-.1,0) -- (.1,0);
  \draw (0,-.1) -- (0,.1);

  \draw (-1.1,-.6) \Triangle{.1};
  \draw (1,-.5) \Square{.1};

  \draw (0,-.1) -- (-1,-.4);
  \draw (0,-.1) -- (1,-.4);

  \draw [very thin,gray,dashed] (-2,0) -- (2,0);
  \draw [very thin,gray,dashed] (-2,-.5) -- (2,-.5);

  \node at (-1,-.8) {Rep};
  \node at (1,-.8) {SPC};
\end{tikzpicture}

%% file: figures/fer1kL128_1024_860.tikz
\begin{tikzpicture}
  \pgfplotsset{
    label style = {font=\fontsize{9pt}{7.2}\selectfont},
    tick label style = {font=\fontsize{7pt}{7.2}\selectfont}
  }

\begin{axis}[
	scale = 1,
    ymode=log,
    xlabel={$E_b/N_0$ [\text{dB}]}, xlabel style={yshift=0.8em},
    ylabel={FER}, ylabel style={yshift=-0.75em},
    grid=both,
    ymajorgrids=true,
    xmajorgrids=true,
    grid style=dashed,
    width=0.5\columnwidth, height=7cm,
    thick,
    mark size=3,
    legend style={
      anchor={center},
      cells={anchor=west},
      column sep= 2mm,
      font=\fontsize{7pt}{7.2}\selectfont,
    },
    legend to name=perf-legend1kL128_1024_860,
    legend columns=2,
]

\addplot[
    color=blue,
    mark=o,
    thick,
    mark size=3,
]
table {
3 0.3681
3.5 0.0337
4 0.000639125
4.5 3.71452e-06
5 4.05754e-08
};
\addlegendentry{\cite{hashemi_SSCL}}

\addplot[
    color=red,
    mark=triangle,
    thick,
    mark size=3,
]
table {
3 0.37
3.5 0.0379
4 0.000550594
4.5 2.60017e-05
5 1.13644e-06
};
\addlegendentry{\cite{sarkis_list}}

\end{axis}
\end{tikzpicture}

%% file: figures/ber1kL128_1024_860.tikz
\begin{tikzpicture}
  \pgfplotsset{
    label style = {font=\fontsize{9pt}{7.2}\selectfont},
    tick label style = {font=\fontsize{7pt}{7.2}\selectfont},
  }

\begin{axis}[
	scale = 1,
    ymode=log,
    xlabel={$E_b/N_0$ [\text{dB}]}, xlabel style={yshift=0.8em},
    ylabel={BER}, ylabel style={yshift=-0.75em},%
    grid=both,
    ymajorgrids=true,
    xmajorgrids=true,
    width=0.5\columnwidth, height=7.0cm,
    grid style=dashed,
    thick,
    mark size=3,
]

\addplot[
    color=blue,
    mark=o,
    thick,
    mark size=3,
]
table {
3 0.134853
3.5 0.0105408
4 0.000164701
4.5 5.55019e-07
5 4.20463e-09
};
%\addlegendentry{\cite{hashemi_SSCL}}

\addplot[
    color=red,
    mark=triangle,
    thick,
    mark size=3,
]
table {
3 0.134756
3.5 0.0117945
4 0.000122962
4.5 1.83733e-06
5 6.06804e-08
};
%\addlegendentry{\cite{sarkis_list}}

\end{axis}
\end{tikzpicture}

%% file: figures/time-req025.tikz
\begin{tikzpicture}

\begin{axis}[
	scale = 1,
    xmode=log,
    log basis x=2,
    xtick=data,
    xlabel={$L$},
    ylabel={Time steps}, %ylabel style={yshift=-0.85em},
    width=\columnwidth, height=7cm,
    legend pos=north east,
    grid=both,
    ymajorgrids=true,
    xmajorgrids=true,
    grid style=dashed,
    xmin = 2,
    xmax = 512,
    ymin = 0,
    ymax = 1200,
    thick,
    mark size = 3,
    legend style={
      font=\fontsize{7pt}{7.2}\selectfont,
    },
    /pgf/number format/.cd,
        use comma,
        1000 sep={}
]

%\addplot[
%    color=orange,
%]
%table {
%2 2302
%4 2302
%8 2302
%16 2302
%32 2302
%64 2302
%128 2302
%256 2302
%512 2302
%%1024 2302
%};
%%\addlegendentry{SCL}

\addplot[
    color=red,
]
table {
2 593
4 593
8 593
16 593
32 593
64 593
128 593
256 593
512 593
%1024 593
};
%\addlegendentry{SSCL}

\addplot[
    color=brown,
]
table {
2 518
4 518
8 518
16 518
32 518
64 518
128 518
256 518
512 518
%1024 518
};
%\addlegendentry{SSCL-SPC}

\addplot[
    color=black,
    mark=square,
]
table {
2 454
4 534
8 578
16 592
32 593
64 593
128 593
256 593
512 593
%1024 593
};
%\addlegendentry{Fast-SSCL}

\addplot[
    color=blue,
    mark=o,
]
table {
2 326
4 393
8 454
16 502
32 518
64 518
128 518
256 518
512 518
%1024 518
};
%\addlegendentry{Fast-SSCL-SPC}

\end{axis}
\end{tikzpicture}

%% file: figures/time-req05.tikz
\begin{tikzpicture}

\begin{axis}[
	scale = 1,
    xmode=log,
    log basis x=2,
    xtick=data,
    xlabel={$L$},
    ylabel={Time steps}, %ylabel style={yshift=-0.85em},
    width=\columnwidth, height=7cm,
    legend pos=north east,
    grid=both,
    ymajorgrids=true,
    xmajorgrids=true,
    grid style=dashed,
    xmin = 2,
    xmax = 512,
    ymin = 0,
    ymax = 1200,
    thick,
    mark size = 3,
    legend style={
      anchor={center},
      cells={anchor=west},
      column sep= 1.5mm,
      font=\fontsize{9pt}{7.2}\selectfont,
    },
    legend to name=perf-legendTimeReq,
    legend columns=5,
    /pgf/number format/.cd,
        use comma,
        1000 sep={}
]

%\addplot[
%    color=orange,
%]
%table {
%2 2558
%4 2558
%8 2558
%16 2558
%32 2558
%64 2558
%128 2558
%256 2558
%512 2558
%%1024 2558
%};
%\addlegendentry{SCL}

\addplot[
    color=red,
]
table {
2 843
4 843
8 843
16 843
32 843
64 843
128 843
256 843
512 843
%1024 843
};
\addlegendentry{SSCL}

\addplot[
    color=brown,
]
table {
2 767
4 767
8 767
16 767
32 767
64 767
128 767
512 767
%1024 767
};
\addlegendentry{SSCL-SPC}

\addplot[
    color=black,
    mark=square,
]
table {
2 447
4 550
8 647
16 737
32 807
64 842
128 843
256 843
512 843
%1024 793
};
\addlegendentry{Fast-SSCL}

\addplot[
    color=blue,
    mark=o,
]
table {
2 320
4 399
8 487
16 571
32 638
64 703
128 767
256 767
512 767
%1024 743
};
\addlegendentry{Fast-SSCL-SPC}

\end{axis}
\end{tikzpicture}

%% file: figures/time-req075.tikz
\begin{tikzpicture}

\begin{axis}[
	scale = 1,
    xmode=log,
    log basis x=2,
    xtick=data,
    xlabel={$L$},
    ylabel={Time steps}, %ylabel style={yshift=-0.85em},
    width=\columnwidth, height=7cm,
    legend pos=north east,
    grid=both,
    ymajorgrids=true,
    xmajorgrids=true,
    grid style=dashed,
    xmin = 2,
    xmax = 512,
    ymin = 0,
    ymax = 1200,
    thick,
    mark size = 3,
    legend style={
      font=\fontsize{7pt}{7.2}\selectfont,
    },
    /pgf/number format/.cd,
        use comma,
        1000 sep={}
]

%\addplot[
%    color=orange,
%]
%table {
%2 2814
%4 2814
%8 2814
%16 2814
%32 2814
%64 2814
%128 2814
%256 2814
%512 2814
%%1024 2814
%};
%%\addlegendentry{SCL}

\addplot[
    color=red,
]
table {
2 1039
4 1039
8 1039
16 1039
32 1039
64 1039
128 1039
256 1039
512 1039
%1024 1039
};
%\addlegendentry{SSCL}

\addplot[
    color=brown,
]
table {
2 972
4 972
8 972
16 972
32 972
64 972
128 972
256 972
512 972
%1024 972
};
%\addlegendentry{SSCL-SPC}

\addplot[
    color=black,
    mark=square,
]
table {
2 372
4 473
8 583
16 705
32 839
64 971
128 1038
256 1039
512 1039
%1024 1039
};
%\addlegendentry{Fast-SSCL}

\addplot[
    color=blue,
    mark=o,
]
table {
2 261
4 337
8 432
16 548
32 679
64 778
128 844
256 972
512 972
%1024 972
};
%\addlegendentry{Fast-SSCL-SPC}

\end{axis}
\end{tikzpicture}

%% file: figures/fer1kL2.tikz
\begin{tikzpicture}
  \pgfplotsset{
    label style = {font=\fontsize{9pt}{7.2}\selectfont},
    tick label style = {font=\fontsize{7pt}{7.2}\selectfont}
  }

\begin{axis}[
	scale = 1,
    ymode=log,
    xlabel={$E_b/N_0$ [\text{dB}]}, xlabel style={yshift=0.8em},
    ylabel={FER}, ylabel style={yshift=-0.75em},
    grid=both,
    ymajorgrids=true,
    xmajorgrids=true,
    grid style=dashed,
    width=0.5\columnwidth, height=7cm,
    thick,
    mark size=3,
    legend style={
      anchor={center},
      cells={anchor=west},
      column sep= 2mm,
      font=\fontsize{7pt}{7.2}\selectfont,
    },
    legend to name=perf-legend1kL2,
    legend columns=4,
]

\addplot[
    color=black,
    mark=square,
    thick,
    mark size=3,
]
table {
1 0.712
1.5 0.3336
2 0.0896
2.5 0.0177
3 0.0031002
%3.5 0.000367985
};
\addlegendentry{$S_{\text{Rate-1}}=0$}

\addplot[
    color=blue,
    mark=o,
    thick,
    mark size=3,
]
table {
1 0.6327
1.5 0.217
2 0.0337
2.5 0.00274605
3 0.000185364
};
\addlegendentry{$S_{\text{Rate-1}}=1$}

\end{axis}
\end{tikzpicture}

%% file: figures/ber1kL2.tikz
\begin{tikzpicture}
  \pgfplotsset{
    label style = {font=\fontsize{9pt}{7.2}\selectfont},
    tick label style = {font=\fontsize{7pt}{7.2}\selectfont},
  }

\begin{axis}[
	scale = 1,
    ymode=log,
    xlabel={$E_b/N_0$ [\text{dB}]}, xlabel style={yshift=0.8em},
    ylabel={BER}, ylabel style={yshift=-0.75em},%
    grid=both,
    ymajorgrids=true,
    xmajorgrids=true,
    width=0.5\columnwidth, height=7.0cm,
    grid style=dashed,
    thick,
    mark size=3,
]

\addplot[
    color=black,
    mark=square,
    thick,
    mark size=3,
]
table {
1 0.22745
1.5 0.0772369
2 0.0142947
2.5 0.00177637
3 0.000267029
%3.5 2.4609e-5
};
%\addlegendentry{$S_{\text{Rate-1}}=0$}

\addplot[
    color=blue,
    mark=o,
    thick,
    mark size=3,
]
table {
1 0.209865
1.5 0.0565697
2 0.00692422
2.5 0.000408045
3 1.83916e-5
};
%\addlegendentry{$S_{\text{Rate-1}}=1$}

\end{axis}
\end{tikzpicture}

%% file: figures/fer1kL4.tikz
\begin{tikzpicture}
  \pgfplotsset{
    label style = {font=\fontsize{9pt}{7.2}\selectfont},
    tick label style = {font=\fontsize{7pt}{7.2}\selectfont}
  }

\begin{axis}[
	scale = 1,
    ymode=log,
    xlabel={$E_b/N_0$ [\text{dB}]}, xlabel style={yshift=0.8em},
    ylabel={FER}, ylabel style={yshift=-0.75em},
    grid=both,
    ymajorgrids=true,
    xmajorgrids=true,
    grid style=dashed,
    width=0.5\columnwidth, height=7cm,
    thick,
    mark size=3,
    legend style={
      anchor={center},
      cells={anchor=west},
      column sep= 2mm,
      font=\fontsize{7pt}{7.2}\selectfont,
    },
    legend to name=perf-legend1kL4,
    legend columns=4,
]

\addplot[
    color=black,
    mark=square,
    thick,
    mark size=3,
]
table {
1 0.6645
1.5 0.293
2 0.0792
2.5 0.0188
3 0.00257719
%3.5 0.000374714
};
\addlegendentry{$S_{\text{Rate-1}}=0$}

\addplot[
    color=red,
    mark=star,
    thick,
    mark size=3,
]
table {
1 0.4544
1.5 0.1026
2 0.0088
2.5 0.000422597
3 1.58083e-5
%3.5 3.24149e-7
};
\addlegendentry{$S_{\text{Rate-1}}=1$}

\addplot[
    color=blue,
    mark=o,
    thick,
    mark size=3,
]
table {
1 0.4557
1.5 0.1027
2 0.0096
2.5 0.000421998
3 1.87398e-5
%3.5 5.06073e-7
};
\addlegendentry{$S_{\text{Rate-1}}=3$}

\end{axis}
\end{tikzpicture}

%% file: figures/ber1kL4.tikz
\begin{tikzpicture}
  \pgfplotsset{
    label style = {font=\fontsize{9pt}{7.2}\selectfont},
    tick label style = {font=\fontsize{7pt}{7.2}\selectfont},
  }

\begin{axis}[
	scale = 1,
    ymode=log,
    xlabel={$E_b/N_0$ [\text{dB}]}, xlabel style={yshift=0.8em},
    ylabel={BER}, ylabel style={yshift=-0.75em},%
    grid=both,
    ymajorgrids=true,
    xmajorgrids=true,
    width=0.5\columnwidth, height=7.0cm,
    grid style=dashed,
    thick,
    mark size=3,
]

\addplot[
    color=black,
    mark=square,
    thick,
    mark size=3,
]
table {
1 0.196899
1.5 0.0609734
2 0.0121152
2.5 0.00233047
3 0.000221175
%3.5 3.33584e-5
};
%\addlegendentry{$S_{\text{Rate-1}}=0$}

\addplot[
    color=red,
    mark=star,
    thick,
    mark size=3,
]
table {
1 0.144138
1.5 0.0259607
2 0.00160371
2.5 6.71203e-5
3 1.6469e-6
%3.5 3.44034e-8
};
%\addlegendentry{$S_{\text{Rate-1}}=1$}

\addplot[
    color=blue,
    mark=o,
    thick,
    mark size=3,
]
table {
1 0.143448
1.5 0.0261775
2 0.00193379
2.5 6.35634e-5
3 1.54383e-6
%3.5 4.18763e-8
};
%\addlegendentry{$S_{\text{Rate-1}}=3$}

\end{axis}
\end{tikzpicture}

%% file: figures/fer1kL8.tikz
\begin{tikzpicture}
  \pgfplotsset{
    label style = {font=\fontsize{9pt}{7.2}\selectfont},
    tick label style = {font=\fontsize{7pt}{7.2}\selectfont}
  }

\begin{axis}[
	scale = 1,
    ymode=log,
    xlabel={$E_b/N_0$ [\text{dB}]}, xlabel style={yshift=0.8em},
    ylabel={FER}, ylabel style={yshift=-0.75em},
    grid=both,
    ymajorgrids=true,
    xmajorgrids=true,
    grid style=dashed,
    width=0.5\columnwidth, height=7cm,
    thick,
    mark size=3,
    legend style={
      anchor={center},
      cells={anchor=west},
      column sep= 2mm,
      font=\fontsize{7pt}{7.2}\selectfont,
    },
    legend to name=perf-legend1kL8,
    legend columns=2,
]

\addplot[
    color=black,
    mark=square,
    thick,
    mark size=3,
]
table {
1 0.6342
1.5 0.2824
2 0.0818
2.5 0.0197
3 0.00314426
};
\addlegendentry{$S_{\text{Rate-1}}=0$}

\addplot[
    color=red,
    mark=star,
    thick,
    mark size=3,
]
table {
1 0.3454
1.5 0.0551
2 0.0028767
2.5 0.000128546
3 3.87097e-6
};
\addlegendentry{$S_{\text{Rate-1}}=1$}

\addplot[
    color=cyan,
    mark=triangle,
    thick,
    mark size=3,
]
table {
1 0.3416
1.5 0.0523
2 0.0032204
2.5 6.51655e-5
3 1.71891e-6
};
\addlegendentry{$S_{\text{Rate-1}}=2$}

\addplot[
    color=blue,
    mark=o,
    thick,
    mark size=3,
]
table {
1 0.3425
1.5 0.051
2 0.00298116
2.5 7.55499e-5
3 2.17841e-6
};
\addlegendentry{$S_{\text{Rate-1}}=7$}

\end{axis}
\end{tikzpicture}

%% file: figures/ber1kL8.tikz
\begin{tikzpicture}
  \pgfplotsset{
    label style = {font=\fontsize{9pt}{7.2}\selectfont},
    tick label style = {font=\fontsize{7pt}{7.2}\selectfont},
  }

\begin{axis}[
	scale = 1,
    ymode=log,
    xlabel={$E_b/N_0$ [\text{dB}]}, xlabel style={yshift=0.8em},
    ylabel={BER}, ylabel style={yshift=-0.75em},%
    grid=both, ymax=4e-1, ymin=7e-8,
    ymajorgrids=true,
    xmajorgrids=true,
    width=0.5\columnwidth, height=7.0cm,
    grid style=dashed,
    thick,
    mark size=3,
]

\addplot[
    color=black,
    mark=square,
    thick,
    mark size=3,
]
table {
1 0.179111
1.5 0.0569701
2 0.0122391
2.5 0.00241387
3 0.000301407
};
%\addlegendentry{$S_{\text{Rate-1}}=0$}

\addplot[
    color=red,
    mark=star,
    thick,
    mark size=3,
]
table {
1 0.107693
1.5 0.0133318
2 0.000483084
2.5 1.54004e-5
3 3.19046e-7
};
%\addlegendentry{$S_{\text{Rate-1}}=1$}

\addplot[
    color=cyan,
    mark=triangle,
    thick,
    mark size=3,
]
table {
1 0.106716
1.5 0.0129973
2 0.000580428
2.5 8.50715e-6
3 1.85694e-7
};
%\addlegendentry{$S_{\text{Rate-1}}=2$}

\addplot[
    color=blue,
    mark=o,
    thick,
    mark size=3,
]
table {
1 0.10661
1.5 0.0131807
2 0.000634195
2.5 1.02199e-5
3 2.45326e-7
};
%\addlegendentry{$S_{\text{Rate-1}}=7$}

\end{axis}
\end{tikzpicture}

%% file: figures/fer1kL4SPC.tikz
\begin{tikzpicture}
  \pgfplotsset{
    label style = {font=\fontsize{9pt}{7.2}\selectfont},
    tick label style = {font=\fontsize{7pt}{7.2}\selectfont}
  }

\begin{axis}[
	scale = 1,
    ymode=log,
    xlabel={$E_b/N_0$ [\text{dB}]}, xlabel style={yshift=0.8em},
    ylabel={FER}, ylabel style={yshift=-0.75em},
    grid=both,
    ymajorgrids=true,
    xmajorgrids=true,
    grid style=dashed,
    width=0.5\columnwidth, height=7cm,
    thick,
    mark size=3,
    legend style={
      anchor={center},
      cells={anchor=west},
      column sep= 2mm,
      font=\fontsize{7pt}{7.2}\selectfont,
    },
    legend to name=perf-legend1kL4SPC,
    legend columns=2,
]

\addplot[
    color=red,
    mark=star,
    thick,
    mark size=3,
]
table {
1 0.4853
1.5 0.1231
2 0.0154
2.5 0.00120552
3 6.04536e-5
};
\addlegendentry{$S_\text{Rate-1}=1$, $S_\text{SPC}=2$}

\addplot[
    color=cyan,
    mark=triangle,
    thick,
    mark size=3,
]
table {
1 0.4712
1.5 0.1037
2 0.0104
2.5 0.000512211
3 2.18925e-5
};
\addlegendentry{$S_\text{Rate-1}=1$, $S_\text{SPC}=3$}

\addplot[
    color=blue,
    mark=o,
    thick,
    mark size=3,
]
table {
1 0.4683
1.5 0.1024
2 0.0083
2.5 0.000412964
3 2.23925e-5
};
\addlegendentry{$S_\text{Rate-1}=3$, $S_\text{SPC}=4$}

\end{axis}
\end{tikzpicture}

%% file: figures/ber1kL4SPC.tikz
\begin{tikzpicture}
  \pgfplotsset{
    label style = {font=\fontsize{9pt}{7.2}\selectfont},
    tick label style = {font=\fontsize{7pt}{7.2}\selectfont},
  }

\begin{axis}[
	scale = 1,
    ymode=log,
    xlabel={$E_b/N_0$ [\text{dB}]}, xlabel style={yshift=0.8em},
    ylabel={BER}, ylabel style={yshift=-0.75em},%
    grid=both,
    ymajorgrids=true,
    xmajorgrids=true,
    width=0.5\columnwidth, height=7.0cm,
    grid style=dashed,
    thick,
    mark size=3,
]

\addplot[
    color=red,
    mark=star,
    thick,
    mark size=3,
]
table {
1 0.154334
1.5 0.0306801
2 0.00306445
2.5 0.000171174
3 7.60156e-6
};
%\addlegendentry{$S_\text{Rate-1}=1$, $S_\text{SPC}=2$}

\addplot[
    color=cyan,
    mark=triangle,
    thick,
    mark size=3,
]
table {
1 0.148879
1.5 0.0266078
2 0.0020002
2.5 7.0369e-5
3 2.52362e-6
};
%\addlegendentry{$S_\text{Rate-1}=1$, $S_\text{SPC}=3$}

\addplot[
    color=blue,
    mark=o,
    thick,
    mark size=3,
]
table {
1 0.148292
1.5 0.0266793
2 0.0016998
2.5 6.28641e-5
3 2.3827e-6
};
%\addlegendentry{$S_\text{Rate-1}=3$, $S_\text{SPC}=4$}

\end{axis}
\end{tikzpicture}

%% file: figures/fer1kL8SPC.tikz
\begin{tikzpicture}
  \pgfplotsset{
    label style = {font=\fontsize{9pt}{7.2}\selectfont},
    tick label style = {font=\fontsize{7pt}{7.2}\selectfont}
  }

\begin{axis}[
	scale = 1,
    ymode=log,
    xlabel={$E_b/N_0$ [\text{dB}]}, xlabel style={yshift=0.8em},
    ylabel={FER}, ylabel style={yshift=-0.75em},
    grid=both,
    ymajorgrids=true,
    xmajorgrids=true,
    grid style=dashed,
    width=0.5\columnwidth, height=7cm,
    thick,
    mark size=3,
    legend style={
      anchor={center},
      cells={anchor=west},
      column sep= 2mm,
      font=\fontsize{7pt}{7.2}\selectfont,
    },
    legend to name=perf-legend1kL8SPC,
    legend columns=2,
]

\addplot[
    color=red,
    mark=star,
    thick,
    mark size=3,
]
table {
1 0.4039
1.5 0.0934
2 0.0093
2.5 0.000855183
3 4.653e-5
};
\addlegendentry{$S_\text{Rate-1}=1$, $S_\text{SPC}=2$}

\addplot[
    color=cyan,
    mark=triangle,
    thick,
    mark size=3,
]
table {
1 0.3536
1.5 0.0587
2 0.0053
2.5 0.000224033
3 9.54946e-6
};
\addlegendentry{$S_\text{Rate-1}=1$, $S_\text{SPC}=3$}

\addplot[
    color=gray,
    mark=asterisk,
    thick,
    mark size=3,
]
table {
1 0.3543
1.5 0.0605
2 0.0051
2.5 0.000163555
3 5.49349e-6
};
\addlegendentry{$S_\text{Rate-1}=1$, $S_\text{SPC}=4$}

\addplot[
    color=green,
    mark=diamond,
    thick,
    mark size=3,
]
table {
1 0.3945
1.5 0.0865
2 0.0095
2.5 0.000942436
3 4.90632e-5
};
\addlegendentry{$S_\text{Rate-1}=2$, $S_\text{SPC}=2$}

\addplot[
    color=brown,
    mark=x,
    thick,
    mark size=3,
]
table {
1 0.3482
1.5 0.0548
2 0.005
2.5 0.000165278
3 7.78845e-6
};
\addlegendentry{$S_\text{Rate-1}=2$, $S_\text{SPC}=3$}

\addplot[
    color=purple,
    mark=square,
    thick,
    mark size=3,
]
table {
1 0.3419
1.5 0.0579
2 0.00324423
2.5 0.000100624
3 2.44589e-6
};
\addlegendentry{$S_\text{Rate-1}=2$, $S_\text{SPC}=4$}

\addplot[
    color=blue,
    mark=o,
    thick,
    mark size=3,
]
table {
1 0.338
1.5 0.0525
2 0.00383318
2.5 0.000102182
3 2.95486e-6
};
\addlegendentry{$S_\text{Rate-1}=7$, $S_\text{SPC}=8$}

\end{axis}
\end{tikzpicture}

%% file: figures/ber1kL8SPC.tikz
\begin{tikzpicture}
  \pgfplotsset{
    label style = {font=\fontsize{9pt}{7.2}\selectfont},
    tick label style = {font=\fontsize{7pt}{7.2}\selectfont},
  }

\begin{axis}[
	scale = 1,
    ymode=log,
    xlabel={$E_b/N_0$ [\text{dB}]}, xlabel style={yshift=0.8em},
    ylabel={BER}, ylabel style={yshift=-0.75em},%
    grid=both,
    ymajorgrids=true,
    xmajorgrids=true,
    width=0.5\columnwidth, height=7.0cm,
    grid style=dashed,
    thick,
    mark size=3,
]

\addplot[
    color=red,
    mark=star,
    thick,
    mark size=3,
]
table {
1 0.12364
1.5 0.0220021
2 0.00180195
2.5 0.000109904
3 4.79841e-6
};
%\addlegendentry{$S_\text{Rate-1}=1$, $S_\text{SPC}=2$}

\addplot[
    color=cyan,
    mark=triangle,
    thick,
    mark size=3,
]
table {
1 0.109228
1.5 0.0140721
2 0.00102051
2.5 2.50462e-5
3 8.24014e-7
};
%\addlegendentry{$S_\text{Rate-1}=1$, $S_\text{SPC}=3$}

\addplot[
    color=gray,
    mark=asterisk,
    thick,
    mark size=3,
]
table {
1 0.110801
1.5 0.0152021
2 0.00075
2.5 2.06168e-5
3 5.37976e-7
};
%\addlegendentry{$S_\text{Rate-1}=1$, $S_\text{SPC}=4$}

\addplot[
    color=green,
    mark=diamond,
    thick,
    mark size=3,
]
table {
1 0.122157
1.5 0.0211902
2 0.00177383
2.5 0.000130468
3 6.05049e-6
};
%\addlegendentry{$S_\text{Rate-1}=2$, $S_\text{SPC}=2$}

\addplot[
    color=brown,
    mark=x,
    thick,
    mark size=3,
]
table {
1 0.108785
1.5 0.0139453
2 0.000978906
2.5 2.55277e-5
3 7.58157e-7
};
%\addlegendentry{$S_\text{Rate-1}=2$, $S_\text{SPC}=3$}

\addplot[
    color=purple,
    mark=square,
    thick,
    mark size=3,
]
table {
1 0.107219
1.5 0.0145793
2 0.000623626
2.5 1.2861e-5
3 2.01595e-7
};
%\addlegendentry{$S_\text{Rate-1}=2$, $S_\text{SPC}=4$}

\addplot[
    color=blue,
    mark=o,
    thick,
    mark size=3,
]
table {
1 0.106937
1.5 0.0128893
2 0.000677095
2.5 1.25533e-5
3 2.7321e-7
};
%\addlegendentry{$S_\text{Rate-1}=7$, $S_\text{SPC}=8$}

\end{axis}
\end{tikzpicture}

%% file: figures/decArch.tikz
\begin{tikzpicture}[scale=1, thick]

\makeatletter
\DeclareRobustCommand{\rvdots}{%
  \vbox{
    \baselineskip4\p@\lineskiplimit\z@
    \kern-\p@
    \hbox{.}\hbox{.}\hbox{.}
  }}
\makeatother

\draw (-1.5,-.75) rectangle (1.5,.25) node [pos=.5,align=center] {$\PM$ Computation \\ and Sorting};

\draw [->] (-.5,.25) -- (-.5,1.1) -- (1,1.1);
\draw [->] (1,.9) -- (.5,.9) -- (.5,.25);

\draw (1,.75) rectangle (3,1.25) node [pos=.5,align=center] {Controller};

\draw [->] (3,1.1) -- (3.75,1.1);
\draw [->] (3.75,.9) -- (3,.9);

\draw (3.75,.75) rectangle (5.75,1.25) node [pos=.5,align=center] {CRC Unit};

\node at (4.75,-.5) {SC Decoders};

\draw (3.5,-.75) rectangle (6,-2.75) node [pos=.5,align=center] {$\cdots$};

\draw [fill=lightgray] (3.6,-.85) rectangle (4.3,-2.65) node [pos=.5,align=center] {\rvdots};

\draw [fill=white] (3.7,-.95) rectangle (4.2,-1.35) node [pos=.5,align=center] {PE};

\draw [fill=white] (3.7,-2.55) rectangle (4.2,-2.15) node [pos=.5,align=center] {PE};

\draw [fill=lightgray] (5.2,-.85) rectangle (5.9,-2.65) node [pos=.5,align=center] {\rvdots};

\draw [fill=white] (5.3,-.95) rectangle (5.8,-1.35) node [pos=.5,align=center] {PE};

\draw [fill=white] (5.3,-2.55) rectangle (5.8,-2.15) node [pos=.5,align=center] {PE};

\node at (3.95,-3) {$1$};
\node at (4.75,-3) {$\cdots$};
\node at (5.55,-3) {$L$};

\node at (6.25,-1.15) {$1$};
\node at (6.25,-1.75) {\rvdots};
\node at (6.25,-2.35) {$P$};

\draw (1,-1.25) rectangle (3,-2.25) node [pos=.5,align=center] {Memories};

\draw [->] (3,-1.5) -- (3.5,-1.5);
\draw [->] (3.5,-2) -- (3,-2);

\draw [->] (.5,-.75) -- (.5,-1.5) -- (1,-1.5);
\draw [->] (1,-2) -- (-.5,-2) -- (-.5,-.75);

\draw [->] (1.75,.75) -- (1.75,-1.25);
\draw [->] (2.25,-1.25) -- (2.25,.75);

\draw [dashed] (-1.75,-3.25) rectangle (6.5,1.5);

\node at (2,2) {Node Sequence};
\draw [->] (2,1.75) -- (2,1.25);

\node at (2,-3.75) {Channel LLRs};
\draw [->] (2,-3.5) -- (2,-2.25);

\end{tikzpicture}

%% file: figures/PEArch.tikz
\begin{tikzpicture}[scale=1, thick]

\draw (1,.5) rectangle (2,1.5) node [pos=.5,align=center] {$\alpha_{i}^{\text{l}}$};

\draw (1,-.5) rectangle (2,-1.5) node [pos=.5,align=center] {$\alpha_{i}^{\text{r}}$};

\draw (3,.5) -- (3.5,.25) -- (3.5,-.25) -- (3,-.5) -- (3,.5);

\draw (2,1) -- (2.5,1) -- (2.5,.25) -- (3,.25);
\draw (2,-1) -- (2.5,-1) -- (2.5,-.25) -- (3,-.25);

\fill (.75,1.25) circle (.05);
\draw (.75,1.25) -- (.75,-.75) -- (1,-.75);

\fill (.5,-1.25) circle (.05);
\draw (.5,-1.25) -- (.5,.75) -- (1,.75);

\draw [dashed] (.25,-1.75) rectangle (3.75,1.75);

\draw (0,2) -- (3.25,2) -- (3.25,.375);
\node at (-.35,2) {$i_s$};

\draw (0,1.25) -- (1,1.25);
\node at (-.25,1.25) {$\alpha_i$};

\draw (0,-1.25) -- (1,-1.25);
\node at (-.5,-1.25) {$\alpha_{i+\frac{N_s}{2}}$};

\draw (3.5,0) -- (4,0);
\node at (4.5,0) {$\alpha_i^{\text{out}}$};

\end{tikzpicture}

%% file: figures/MemoryArch.tikz
\begin{tikzpicture}[scale=0.95, thick]

\draw (.5,.5) rectangle ++(1.5,1.5);

\draw [fill=white] (0,0) rectangle ++(1.5,1.5) node [pos=.5,align=center] {High \\ Stage \\ Memory};

\node at (1.75,1) {\reflectbox{$\ddots$}};

\draw [<->] (1.6,-.1) -- ++(.5,.5) node [midway, below, sloped] {$L$};

\draw [<->] (0,-.2) -- ++(1.5,0) node [midway, below, sloped] {$Q_{\text{LLR}}\times P$};

\draw [<->] (-.2,0) --  ++(0,1.5) node [midway, above, sloped] {$N/P-2$};

\draw (4,.5) rectangle ++(1.5,1.5);

\draw [fill=white] (3.5,0) rectangle ++(1.5,1.5) node [pos=.5,align=center] {Low \\ Stage \\ Memory};

\node at (5.25,1) {\reflectbox{$\ddots$}};

\draw [<->] (5.1,-.1) -- ++(.5,.5) node [midway, below, sloped] {$L$};

\draw [<->] (3.5,-.2) -- ++(1.5,0) node [midway, below, sloped] {$Q_{\text{LLR}}$};

\draw [<->] (3.3,0) --  ++(0,1.5) node [midway, above, sloped] {$2P-2$};

\draw [fill=white] (6.5,0) rectangle ++(1.5,1.5) node [pos=.5,align=center] {Channel \\ Memory};

\draw [<->] (6.5,-.2) -- ++(1.5,0) node [midway, below, sloped] {$Q_{\text{LLR}}$};

\draw [<->] (6.3,0) --  ++(0,1.5) node [midway, above, sloped] {$N$};

\draw [dashed] (-1,-.75) rectangle ++(9.3,3);
\node at (7,2) {LLR Memories};

\draw (.25,-2.75) rectangle ++(1.5,1.5);

\draw [fill=white] (-0.25,-3.25) rectangle ++(1.5,1.5) node [pos=.5,align=center] {Path \\ Memory};

\node at (1.5,-2.25) {\reflectbox{$\ddots$}};

\draw [<->] (1.35,-3.35) -- ++(.5,.5) node [midway, below, sloped] {$L$};

\draw [<->] (-.25,-3.45) -- ++(1.5,0) node [midway, below, sloped] {$1$};

\draw [<->] (-.45,-3.25) --  ++(0,1.5) node [midway, above, sloped] {$N$};

\draw [dashed] (-1,-4) rectangle ++(3.20,3);

\draw (3.5,-2.75) rectangle ++(1.5,1.5);

\draw [fill=white] (3,-3.25) rectangle ++(1.5,1.5) node [pos=.5,align=center] {$\beta$ \\ Memory};

\node at (4.75,-2.25) {\reflectbox{$\ddots$}};

\draw [<->] (4.6,-3.35) -- ++(.5,.5) node [midway, below, sloped] {$L$};

\draw [<->] (3,-3.45) -- ++(1.5,0) node [midway, below, sloped] {$1$};

\draw [<->] (2.8,-3.25) --  ++(0,1.5) node [midway, above, sloped] {$N-1$};

\draw [dashed] (2.25,-4) rectangle ++(3.2,3);

\draw (6.65,-2.75) rectangle ++(1.35,1);

\draw [fill=white] (6.15,-3.25) rectangle ++(1.35,1) node [pos=.5,align=center] {$\PM$ \\ Memory};

\node at (7.75,-2.5) {\reflectbox{$\ddots$}};

\draw [<->] (7.60,-3.35) -- ++(.5,.5) node [midway, below, sloped] {$L$};

\draw [<->] (6.15,-3.45) -- ++(1.35,0) node [midway, below, sloped] {$Q_{\PM}$};

\draw [<->] (5.95,-3.25) --  ++(0,1) node [midway, above, sloped] {$1$};

\draw [dashed] (5.5,-4) rectangle ++(2.8,3);

\end{tikzpicture}

%% file: figures/PathMemMod.tikz
\begin{tikzpicture}[scale=.8, thick]
\footnotesize

% \draw (0,0) rectangle ++(1.25,.5) node [pos=.5,align=center] {$N_v=64$};

\draw (3,-2) rectangle ++(3,-4) node [pos=.5,align=center] {Path \\ Memory}; 
\draw (2,-1) -- ++(2.5,0) -- ++(0,-1);
\draw (2,-7) -- ++(2,0) -- ++(0,1);
\node at (3, -7.4) {Address};
\node at (3, -.6) {Data in};
\draw (5,-8.5) -- ++(0,2.5);
\node at (5, -9) {Write Enable};

\node at (-1.7,-4.1) [align=right]{Node Type};
\draw (-.7,-4.1) -- (1,-4.1);
\draw (1,-3.65) -- ++(0,-0.965);

%%%%%%%%%First MUX

\draw (0,0.65) -- ++(2,-.5) -- ++(0,-3.55) -- ++(-2,-.5) -- ++(0,4.55);
\node at (0.6,-0.1) [align=left]{RATE0};
\node at (0.76,-0.45)[align=left]{RATE1-1};
\node at (0.76,-0.80)[align=left]{RATE1-2};
\node at (0.51,-1.15)[align=left]{REP1};
\node at (0.51,-1.50)[align=left]{REP2};
\node at (0.51,-1.85)[align=left]{SPC1};
\node at (0.62,-2.20)[align=left]{SPC2-1};
\node at (0.62,-2.55)[align=left]{SPC2-2};
\node at (0.48,-2.90)[align=left]{SPC3};
\node at (0.55,-3.25)[align=left]{LEAF};

\draw (-.5,-.1) -- ++(.5,0);
\draw (-.5,-.45) -- ++(.5,0);
\draw (-.5,-.8) -- ++(.5,0);
\draw (-.5,-1.15) -- ++(.5,0);
\draw (-.5,-1.5) -- ++(.5,0);
\draw (-.5,-1.85) -- ++(.5,0);
\draw (-.5,-2.20) -- ++(.5,0);
\draw (-.5,-2.55) -- ++(.5,0);
\draw (-.5,-2.9) -- ++(.5,0);
\draw (-.5,-3.25) -- ++(.5,0);

\node at (-.7,-0.1) [align=right]{$0$};
\node at (-.7,-0.45)[align=right]{$\hat{u}$};
\node at (-1.1,-0.80)[align=right]{$\sgn(\alpha)$};
\node at (-.7,-1.15)[align=right]{$0$};
\node at (-.7,-1.50)[align=right]{$\hat{u}$};
\node at (-.7,-1.85)[align=right]{$0$};
\node at (-.7,-2.20)[align=right]{$\hat{u}$};
\node at (-1.1,-2.55)[align=right]{$\sgn(\alpha)$};
\node at (-1,-2.90)[align=right]{$\beta_{i_{\min}}$};
\node at (-.7,-3.25)[align=right]{$\hat{u}$};

%%%%Second MUX

\draw (0,-4.35) -- ++(2,-.5) -- ++(0,-3.55) -- ++(-2,-.5) -- ++(0,4.55);
\node at (0.6,-5.1) [align=left]{RATE0};
\node at (0.76,-5.45)[align=left]{RATE1-1};
\node at (0.76,-5.80)[align=left]{RATE1-2};
\node at (0.51,-6.15)[align=left]{REP1};
\node at (0.51,-6.50)[align=left]{REP2};
\node at (0.51,-6.85)[align=left]{SPC1};
\node at (0.62,-7.20)[align=left]{SPC2-1};
\node at (0.62,-7.55)[align=left]{SPC2-2};
\node at (0.48,-7.90)[align=left]{SPC3};
\node at (0.55,-8.25)[align=left]{LEAF};

\draw (-.5,-5.1) -- ++(.5,0);
\draw (-.5,-5.45) -- ++(.5,0);
\draw (-.5,-5.8) -- ++(.5,0);
\draw (-.5,-6.15) -- ++(.5,0);
\draw (-.5,-6.5) -- ++(.5,0);
\draw (-.5,-6.85) -- ++(.5,0);
\draw (-.5,-7.20) -- ++(.5,0);
\draw (-.5,-7.55) -- ++(.5,0);
\draw (-.5,-7.9) -- ++(.5,0);
\draw (-.5,-8.25) -- ++(.5,0);

\node at (-1.3,-5.1) [align=right]{$i\rightarrow i+2^s$};
\node at (-.7,-5.45)[align=right]{$i$};
\node at (-1.96,-5.80)[align=right]{$i\rightarrow i+2^s-S_\text{Rate-1}$};
\node at (-1.60,-6.15)[align=right]{$i\rightarrow i+2^s-1$};
\node at (-.7,-6.50)[align=right]{$i$};
\node at (-.7,-6.85)[align=right]{$i$};
\node at (-.7,-7.20)[align=right]{$i$};
\node at (-2.16,-7.55)[align=right]{$i\rightarrow i+2^s-S_\text{SPC}-1$};
\node at (-.9,-7.90)[align=right]{$i_{\min}$};
\node at (-.7,-8.25)[align=right]{$i$};

\end{tikzpicture}

%% file: figures/CRCArch.tikz
\begin{tikzpicture}[scale=0.98, thick]
\footnotesize
\draw (0,0) rectangle ++(1.5,.5) node [pos=.5,align=center] {$N_{\text{CRC}}=64$};

\draw (0,-.75) rectangle ++(1.5,.5) node [pos=.5,align=center] {$N_{\text{CRC}}=63$};

\draw (0,-1.5) rectangle ++(1.5,.5) node [pos=.5,align=center] {$N_{\text{CRC}}=32$};

\draw (0,-2.25) rectangle ++(1.5,.5) node [pos=.5,align=center] {$N_{\text{CRC}}=31$};

\draw (0,-3) rectangle ++(1.5,.5) node [pos=.5,align=center] {$N_{\text{CRC}}=16$};

\draw (0,-3.75) rectangle ++(1.5,.5) node [pos=.5,align=center] {$N_{\text{CRC}}=15$};

\draw (0,-4.5) rectangle ++(1.5,.5) node [pos=.5,align=center] {$N_{\text{CRC}}=8$};

\draw (0,-5.25) rectangle ++(1.5,.5) node [pos=.5,align=center] {$N_{\text{CRC}}=7$};

\draw (0,-6) rectangle ++(1.5,.5) node [pos=.5,align=center] {$N_{\text{CRC}}=4$};

\draw (0,-6.75) rectangle ++(1.5,.5) node [pos=.5,align=center] {$N_{\text{CRC}}=3$};

\draw (0,-7.5) rectangle ++(1.5,.5) node [pos=.5,align=center] {$N_{\text{CRC}}=2$};

\draw (0,-8.25) rectangle ++(1.5,.5) node [pos=.5,align=center] {$N_{\text{CRC}}=1$};

\draw (4,-3.75) -- ++(0,1.75) -- ++(1,-.5) -- ++(0,-2.25) -- ++(-1,-.5) -- ++(0,1.5);

\draw (1.5,.25) -- ++(2,0) -- ++(0,-2.5) -- ++(.5,0);

\draw (1.5,-.5) -- ++(1.75,0) -- ++(0,-2) -- ++(.75,0);

\draw (1.5,-1.25) -- ++(1.5,0) -- ++(0,-1.5) -- ++(1,0);

\draw (1.5,-2) -- ++(1.25,0) -- ++(0,-1) -- ++(1.25,0);

\draw (1.5,-2.75) -- ++(1,0) -- ++(0,-.5) -- ++(1.5,0);

\draw (1.5,-3.5) -- ++(.75,0) -- ++(0,0) -- ++(1.75,0);

\draw (1.5,-4.25) -- ++(1,0) -- ++(0,.5) -- ++(1.5,0);

\draw (1.5,-5) -- ++(1.25,0) -- ++(0,1) -- ++(1.25,0);

\draw (1.5,-5.75) -- ++(1.5,0) -- ++(0,1.5) -- ++(1,0);

\draw (1.5,-6.5) -- ++(1.75,0) -- ++(0,2) -- ++(0.75,0);

\draw (1.5,-7.25) -- ++(2,0) -- ++(0,2.5) -- ++(.5,0);

\draw (1.5,-8) -- ++(2.25,0) -- ++(0,3) -- ++(.25,0);

\draw (-1,-8.5) -- ++(1.75,0) -- ++(0,.25);

\draw [dashed] (-.75,-8.75) rectangle ++(6,9.5);

\draw (-.5,-8) -- ++(.5,0);

\draw (-.5,-3.5) -- ++(0,3.75);
\draw (-.5,-3.5) -- ++(0,-4.5);

\draw (-.5,.25) -- ++(.5,0);
\draw (-.5,-.5) -- ++(.5,0);
\draw (-.5,-1.25) -- ++(.5,0);
\draw (-.5,-2) -- ++(.5,0);
\draw (-.5,-2.75) -- ++(.5,0);
\draw (-1,-3.5) -- ++(1,0);
\draw (-.5,-4.25) -- ++(.5,0);
\draw (-.5,-5) -- ++(.5,0);
\draw (-.5,-5.75) -- ++(.5,0);
\draw (-.5,-6.5) -- ++(.5,0);
\draw (-.5,-7.25) -- ++(.5,0);

\node at (-1.75,-8.25) {Estimated};
\node at (-1.75,-8.65) {bit};

\draw (5,-3.75) -- ++(.5,0);

\node at (-1.75,-3.5) {Remainder};

\draw (-1,1) -- ++(5.5,0) -- ++(0,-3.25);

\node at (-1.75,1.25) {Node Type};
\node at (-1.75,.75) {Node Size};

\end{tikzpicture}

%% file: figures/AreaLatencyL2.tikz
\begin{tikzpicture}[scale=1]
\scriptsize
  \pgfplotsset{
    label style = {font=\fontsize{9pt}{7.2}\selectfont},
    tick label style = {font=\fontsize{7pt}{7.2}\selectfont}
  }

\begin{axis}[
	scale = 1,
    xlabel={Latency [$\upmu$s]},
    ylabel={Area [mm$^2$]},
    xmode=log,
    ymode=log,
    grid=both,
    ymajorgrids=true,
    xmajorgrids=true,
    grid style=dashed,
    thick,
    mark size=3,
]

\addplot[
    color=black,
    mark=square,
    thick,
    mark size=2,
    nodes near coords,
	only marks,
	point meta=explicit symbolic,
	visualization depends on={value \thisrow{anchor}\as\myanchor},
	every node near coord/.append style={anchor=\myanchor}
]
table [meta=label] {
x y label anchor
0.55 1.048 {This work} south
0.57 1.03 \cite{lin_SCL} north
0.83 0.68 \cite{hashemi_SSCL_TCASI} south
};
\addlegendentry{\normalsize $L=2$}

\addplot[
    color=blue,
    mark=triangle,
    thick,
    mark size=2,
    nodes near coords,
	only marks,
	point meta=explicit symbolic,
	visualization depends on={value \thisrow{anchor}\as\myanchor},
	every node near coord/.append style={anchor=\myanchor}
]
table [meta=label] {
x y label anchor
0.64 1.822 {This work} north
1.10 0.62 \cite{yuan_multibit_LLR} south
1.86 0.73 \cite{xiong_symbol} south
0.66 2.00 \cite{lin_SCL} south
0.69 0.99 \cite{xiong_multimode} north
0.89 1.22 \cite{hashemi_SSCL_TCASI} south
};
\addlegendentry{\normalsize $L=4$}

\addplot[
    color=red,
    mark=o,
    thick,
    mark size=2,
    nodes near coords,
	only marks,
	point meta=explicit symbolic,
	visualization depends on={value \thisrow{anchor}\as\myanchor},
	every node near coord/.append style={anchor=\myanchor}
]
table [meta=label] {
x y label anchor
0.85 3.975 {This work} south
1.19 3.11 \cite{hashemi_SSCL_TCASI} south
};
\addlegendentry{\normalsize $L=8$}

\end{axis}
\end{tikzpicture}